\documentclass[graybox]{svmult}

\usepackage{mathptmx}       
\usepackage{helvet}         
\usepackage{courier}        
\usepackage{type1cm}        
\usepackage{makeidx}         
\usepackage{graphicx}        
\usepackage{multicol}        
\usepackage[bottom]{footmisc}
\usepackage{amsmath,amssymb}

\newcommand{\yl}[1]{\textcolor{black}{#1}}

\makeindex             

\newcommand{\bB}{\bm{B}}
\newcommand{\bU}{\bm{U}}

\newcommand{\bbb}{\beta}

\newcommand{\bgam}{\bm{\gamma}}
\newcommand{\btheta}{\bm{\theta}}
\newcommand{\bbeta}{\bm{\beta}}

\newcommand{\be}{\bm{e}}

\newcommand{\tbgam}{\tilde{\bgam}}
\newcommand{\hbgam}{\hat{\bgam}}

\newcommand{\htheta}{\hat{\theta}}
\newcommand{\ttheta}{\tilde{\theta}}

\newcommand{\hbeta}{\hat{\bbb}}

\newcommand{\hbbeta}{\hat{\bbeta}}

\newcommand{\sigl}{\sigma_{nj}}
\newcommand{\hsigl}{\hat{\sigma}_{nj}}

\newcommand{\mS}{\mathbb{S}}
\newcommand{\mI}{\mathcal{I}}

\newcommand{\mF}{\mathbb{F}}

\newcommand{\mH}{\mathbb{H}}
\newcommand{\mR}{\mathbb{R}}

\newcommand{\cR}{\mathcal{R}}

\newcommand{\argmax}[1]{\underset{#1}{\rm arg\, max\,\,}}

\newcommand{\ba}{\bm{a}}

\newcommand{\bz}{\bm{z}}
\newcommand{\bZ}{\bm{Z}}

\newcommand{\hsig}{\hat{\sigma}}

\usepackage{bm} 
\usepackage[shortlabels]{enumitem} 
\setlist[enumerate]{wide=0pt, leftmargin=40pt, labelwidth=6pt, align=left}
\usepackage{graphicx,caption}
\usepackage[square,sort,comma,numbers]{natbib}
\usepackage{longtable,booktabs}
\usepackage{setspace}

\usepackage[flushleft]{threeparttable}

\begin{document}

\title*{A Soft-Thresholding Operator for Sparse Time-Varying Effects in  Survival Models}
\author{Yuan Yang, Jian Kang and Yi Li}
\institute{Yuan Yang\at Parexel, Waltham: {Anna.Yang@parexel.com}
\and Jian Kang\at University of Michigan, Ann Arbor: {jiankang@umich.edu}
\and Yi Li \at University of Michigan, Ann Arbor: {yili@umich.edu}}
%
%
\maketitle


\abstract{We consider a class of Cox models with time-dependent effects that may be zero over certain unknown time regions or, in short, sparse time-varying effects. The model is particularly useful for biomedical studies as it conveniently depicts the gradual evolution of effects of risk factors on survival. Statistically, estimating and drawing inference on infinite dimensional functional parameters with sparsity (e.g., time-varying effects with zero-effect time intervals) present enormous challenges. 
To address them, we propose a new soft-thresholding operator for modeling sparse, piecewise smooth and continuous time-varying coefficients in a Cox time-varying effects model.   
Unlike the common regularized methods,  our approach enables one to estimate non-zero time-varying effects  and detect zero regions  simultaneously, and construct a new type of sparse confidence intervals that accommodate zero regions. This leads to a more interpretable model with a straightforward inference procedure. We develop an efficient algorithm for inference in the target functional space,  show that the proposed method enjoys desired theoretical properties, and
present its finite sample performance by way of simulations. We apply the proposed method to analyze the data of the Boston Lung Cancer Survivor  Cohort, an epidemiological cohort study investigating the impacts of risk factors on lung cancer survival, and obtain  clinically useful results.
}

\section{Introduction}
The Cox proportional hazards model, proposed by the late Sir D.R. Cox  \cite{cox1972regression}, has dominated the analysis of survival studies for decades and has also become an indispensable analytical  tool in the era of precision medicine, because of its elegant estimation and inference framework
and ease of interpretation \citep{hong2019quantile}. One key assumption of the proportional hazards model is that the effect of a given covariate remains constant over time, which, however, may not always hold. In fact, non-proportionality has been  commonly observed  and sparked much interest  \cite{Schoenfeld1982,Huang1999,Martinussen2002,Marzec1997,Murphy1993,Winnett2003}, 
which led to the development of  time-dependent coefficients Cox models \cite{Hastie1993}. 

An often overlooked feature in time-dependent effects Cox models is the sparsity  associated with time-varying effects, 
  meaning that the covariate effects can be zero on some specific time intervals and non-zero but time-varying on the others. For example, Anderson and Gill (1982) \cite{Anderson1982} noticed the effects of some covariates disappeared in the later follow-up in a vulvar cancer study; Gore et al. (1982) \cite{Gore1984} found that the influence of signs recorded at diagnosis waned with time in a breast cancer study; Tian et al. (2005) \cite{Tian2005} noted sparsity in the edema effect during the early stage and also showed the effect of prothrombin on survival diminished over time in a biliary cirrhosis study.
  In the motivating  Boston Lung Cancer Survivor Cohort \cite{Christiani2017}, an epidemiological  study investigating  the impacts of clinical and molecular risk factors on lung cancer survival,  chemotherapy and radiotherapy did not seem to increase or decrease patients' overall survival, leading to the notion of detecting zero-effect regions for these treatment options.

It is rather challenging to detect no-effects periods and estimate  non-zero effects simultaneously, as the existing methods for fitting the time-dependent Cox models cannot achieve these goals. For example, the commonly used penalized spline  models \cite{Zucker1990,Yan2012,Lian2013,He2017}
and kernel weighted likelihood approaches \cite{CaiSun2003,Tian2005} can detect or label covariates as time-varying or time-constant, but cannot detect  no-effects periods within each covariate effect's trajectory. 

We propose a new statistical method that can efficiently model sparse time-varying effects in a survival setting,  by using a soft-thresholding operator to represent the time-varying effects in the Cox model.  \yl{Both  soft-thresholding and  hard-thresholding approaches can be applied in this setting and have their own merits. However, we opt for soft-thresholding because it respects the continuity of the effect with respect to time  and  may conveniently depict the gradual evolution of effects of risk factors on survival.}
Indeed, the concept of soft thresholding was introduced by Donoho (1994, 1995) \cite{Donoho1994, Donoho1995},  who applied this estimator to the coefficients of a wavelet transform of a function measured with noise.  
Since then,  the use of soft-thresholding for effect shrinkaging has been flourishing: Chang et al. (2000) \cite{ChangYV2000} proposed an adaptive, data-driven thresholding method for image denoising in a Bayesian framework;
 Tibshirani (1996) \cite{Tibshirani1996} pointed out that the Lasso estimator is a soft-thresholding estimator when the covariate matrix has an orthonormal design. Kang et al. (2018) \cite{Kang2018}  used a soft-thresholding operator for modeling sparse, continuous, and piecewise smooth functions in image data analysis; however, as their method was not developed for survival analysis, it is unclear whether its extension to a survival setting is feasible.

We propose a soft-thresholding operator to model time-varying effects of covariates in a survival regression setting, and use the B-splines to approximate the nonparametric parts. Estimation is carried out by maximizing a penalized and smoothed partial likelihood. We prove the asymptotic properties of our proposed estimator, and introduce a new class of sparse confidence intervals for quantifying the uncertainty of the sparse functional estimates.

This chapter is organized as follows. In Section ~\ref{ch3:sec:methods}, we introduce the proposed
 soft-thresholding operator for a Cox model with sparse time-varying effects and derive an algorithm for fitting the model. Section \ref{ch3:sec:asymp} lists the theoretical properties of 
 the method and  proposes the sparse confidence intervals for inferring from the estimated sparse time-varying effects. We present simulation results in Section \ref{ch3:sec:simul} to assess the finite sample performance of our methods, and analyze the aforementioned Boston Lung Cancer study in Section \ref{ch3:sec:real}. Section \ref{ch3:sec:discussion} concludes this chapter with a brief summary. We defer all the technical proofs to the Appendix.

\section{Methods}\label{ch3:sec:methods}

\subsection{Model}
Let $T^u_i$ and $T^c_i$ represent the survival and censoring times, respectively, for the $i$th patient. Observed are  $T_i=T^u_i \wedge T^c_i$ with $a \wedge b = \min\{a,b\}$, and  the death indicators  $\Delta_i= \mI(T^u_i\le T^c_i)$ with $\mI(A)$  indicating whether condition $A$ holds
($=1$) or not ($=0$). Let $\bm{Z}_i=(Z_{i1},\dots, Z_{ip})^\top$ be a $p$-dimensional covariate vector for sample $i$. 
The observed data consist of $n$ independent vectors, $(T_i, \Delta_i, \bm{Z}_i)$, which 
are identical and independently distributed (i.i.d.) copies of $(T, \Delta, \bm{Z})$. Further, $(T^u_i, T^c_i), i=1, \ldots, n,$ are i.i.d. copies
of $(T^u, T^c)$.

Denote by $\lambda(t|\bm{Z}_i)$ the hazard function at $t$ given $\bm{Z}_i$. A time-varying effects Cox model stipulates that
\begin{align*}
	\lambda(t\mid \bm{Z}_i)=\lambda_{0}(t)\exp\{\bm{Z}_i^\top\bm{\beta}(t)\},
\end{align*}
where $\lambda_0(t)$ is the baseline hazard, and $\bm{\beta}(t) = \{\beta_1 (t), 
\ldots, \beta_p(t)\}^{\top}$ are the $p$ time-dependent coefficients corresponding to $\bm{Z}_i$.

The log partial likelihood with time-varying coefficients is
\begin{align}\label{ch3:eq:PL0}
{\rm PL}(\bbeta) = \sum_{i=1}^{n} \Delta_i \left\{ \sum_{j=1}^p Z_{ij}\beta_j(T_i) -\log\left[ \sum_{l\in R_i} \exp\left\{ \sum_{j=1}^p Z_{lj}\beta_j(T_i) \right\}   \right] \right\},
\end{align}
where $R_i= \{l: T_l>T_i\}$ is the risk set at  $T_i$.

We assume that $\beta_j (\cdot), j=1, \ldots, p,$ is continuous everywhere, with  zero-effect regions ($\cR_{0}$) consisting of at least one interval, and  is smooth over regions (positive $\cR_{+}$ and negative $\cR_{-}$) where its effect is non-zero. On each interval of the non-zero regions, the $d$th derivative of $\beta_j(t)$  exists and satisfies the Lipschitz condition. That is, for any $s, t$ in the interval, there exists a constant $C>0$ such that
\begin{align}\label{ch3:defn:Lipschitz}
|\beta_j^{(d)}(s)-\beta_j^{(d)}(t)| \leq C|s-t|^{w}, 
\end{align}
where $d$ is a non-negative integer, and $w \in (0,1]$ such that $m \equiv d+w> \mbox{0$\cdot$5}$.  Let $\mH$ be the set of all such functions. Often, we use $d=3$ (as in our later simulations and data analysis) corresponding to piecewise cubic functions.  
Let  $\bbeta_0(\cdot)=\{\beta_{01}(\cdot),\ldots,\beta_{0p}(\cdot)\}^\top$ be the true coefficient vector in the model that generates the observed data and $\beta_{0j}\in \mH$.

We use the soft-thresholding operator $\zeta$ to represent a varying coefficient with zero regions:
\begin{align*}
	\zeta{\left\{\theta(t),\alpha\right\} }= \left\{\theta(t)-\alpha\right\} \mI\{\theta(t)>\alpha\} + \left\{\theta(t)+\alpha\right\}\mI\{\theta(t)<-\alpha\},
\end{align*}
where $\alpha>0$ is the thresholding parameter and $\theta(t)$ is a real-valued function.

To avoid technicalities at the tail of the distribution of $T^u$, we estimate
$\beta_j(\cdot)$  over a finite interval $(0, \tau)$ and base the estimation
on the partial likelihood over the same interval, where $\tau$ is within the support
of $T$. By doing so, we need to effectively replace $T_i$ and $\Delta_i$ in likelihood (\ref{ch3:eq:PL0}) (and also the modified partial likelihoods thereafter)
by $T_i = \min(T_i^u, T_i^c, \tau)$ and $\Delta_i= I(T_i^ u \le  T_i^c \wedge\tau)$.
In practice, if $\tau$ is chosen to be the maximum observed survival time in the data, no such replacements are needed. 

It can be shown  that for any function $ \bbb(t) \in \mH$ and any $\alpha>0$, there exists at least one  $\theta(t) \in \mF_0$ such that $\bbb(t) = \zeta{\{\theta,\alpha\}}(t)$, where $\mF_0$ is the class of functions $\theta$ defined on $[0,\tau]$, with the $d$th derivative $\theta^{(d)}$ satisfying the Lipschitz condition  \eqref{ch3:defn:Lipschitz}. 
As such, we introduce a new penalized likelihood for estimation
\begin{eqnarray}\label{ch3:eq:PL1}
& & {\rm PL}(\btheta)   \nonumber \\
& = & \sum_{i=1}^{n} \Delta_i \left\{ \sum_{j=1}^p Z_{ij}  \zeta\{\theta_j(T_i),\alpha_j\}    -\log\left[ \sum_{l\in R_i} \exp\left\{ \sum_{j=1}^p Z_{lj}\zeta\{\theta_j(T_l),\alpha_j\} \right\}   \right] \right\} -\rho ||\btheta||_2^2, \nonumber \\
\end{eqnarray}
where $\btheta(t) = \{\theta_1(t),\ldots,\theta_p(t)\}^\top$ and $\rho >0$ is the predetermined penalization coefficient.

With the soft-thresholding representation, we can convert the problem from estimating non-smooth functions to estimating smooth functions. Among many approaches for modeling smooth functions, we will utilize the B-spline basis approach because of its convenience and numerical stability
\citep{wahba1980ill};   \yl{ other alternatives may include 
 P-splines and smoothing splines. } Let $\mF$ be the B-spline function sieve space, 
 $K = O(n^{\nu})$ be an integer with $0< \nu<\mbox{0$\cdot$5}$, and  $B_k(t) (1\leq k \leq q, \text{ and } q=K +d)$ be the B-spline basis functions of degree $d+1$ associated with the knots $0 = t_0 < t_1 < \dots < t_{K-1} < t_{K} = 1$, satisfying $\max_{1 \le k \le K} (t_k - t_{k-1})=O(n^{-\nu})$. Let $\bB(t) =\{B_{1}(t),\dots, B_{q}(t)\}^\top$ be a functional vector of the B-spline bases; with  $d=3$, this  corresponds to a vector of cubic B-spline bases. 
 Then, we have
\begin{align*}
	\mF = \left\{ \theta(t): \theta(t) = \sum_{k=1}^{q} \gamma_k B_k(t), t \in [0,\tau], \gamma_k \in \mR, k=1,\dots, q \right\}.
\end{align*}

For given $\alpha$ and $q$, we define the thresholding sieve space $$\mS_{q,\alpha} = \left\{ \bbb(t) = \zeta{\left\{\theta(t),\alpha \right\}}: \theta(t) = \sum_{k=1}^{q} \gamma_k B_k(t), t \in [0,\tau], \gamma_k \in \mR, k=1,\dots, q \right\}.$$ 
Let $\bgam_j = (\gamma_{j1},\ldots, \gamma_{jq})^\top$ be the basis coefficients for $\theta_j(t)$. Then we represent $\theta_j(t) = \bB(t)^\top\bgam_j$.  The penalized log partial likelihood can be written as
\begin{align}\label{ch3:eq:PL2}
{\rm PL}(\bgam) = & \sum_{i=1}^{n} \Delta_i \left\{ \sum_{j=1}^p Z_{ij}  \zeta\{\bB(T_i)^\top\bgam_j,\alpha_j\}    -\log\left[ \sum_{l\in R_i} \exp\left\{ \sum_{j=1}^p Z_{lj}\zeta\{\bB(T_i)^\top\bgam_j,\alpha_j\} \right\}   \right] \right\} \nonumber\\
&-\rho \sum_{j=1}^{p} \sum_{i=1}^{n} \{\bB(T_i)^\top\bgam_j\}^2,
\end{align}
where $\bgam = (\bgam_1,\ldots,\bgam_p)$.

\subsection{Estimation}

It is challenging to directly maximize the likelihood function (\ref{ch3:eq:PL2})
as the thresholding operator $\zeta{(\theta, \alpha)}$ is non-smooth. We therefore consider a smooth approximation of  $\zeta{(\theta, \alpha)}$:
\begin{align*}
	h_\eta\{\theta(t), \alpha\}
	=& \frac{1}{2}\Bigg(\left[ 1+\frac{2}{\pi} {\rm arctan}\{\theta_-(t)/\eta\} \right]\theta_-(t) +\\
	&\left[ 1-\frac{2}{\pi} {\rm arctan}\{\theta_+(t)/\eta\}\right] \theta_+(t)\Bigg),    
\end{align*}
where $\alpha >0$,  $\eta >0$ and $\theta_{\pm}(t) = \theta(t)\pm \alpha$.  
\yl{ Noting $ \lim_{\eta \rightarrow 0} h_{\eta}\{\theta(t),\alpha\} =\xi(\theta, \alpha)$, 
 we define $h_{0}\{\theta(t),\alpha\} =\xi(\theta, \alpha)$. As such,
$h_{\eta}\{\theta(t),\alpha\}$ is a sufficiently smooth function in $\eta$ and, in particular,  in a small neighborhood, e.g., $\eta \in [0, \epsilon)$ where  $\epsilon>0$ is small. Taking a Taylor expansion of  $h_{\eta}\{\theta(t),\alpha\}$ at $\eta=0$ within this neighborhood, we can show that } the approximation error between $h_\eta\{\theta(t), \alpha\}$ and $\zeta{(\theta, \alpha)}$ is bounded by $\eta+ O({\eta}^3)$. We drop $\eta$ hereafter for simplicity of notation. Then, we obtain a smoothed log partial likelihood function:
\begin{align}\label{ch3:eq:PL3}
{\rm PL}(\bgam) = & \sum_{i=1}^{n} \Delta_i \left\{ \sum_{j=1}^p Z_{ij}  h\{\bB(T_i)^\top\bgam_j,\alpha_j\}    -\log\left[ \sum_{l\in R_i} \exp\left\{ \sum_{j=1}^p Z_{lj}h\{\bB(T_i)^\top\bgam_j,\alpha_j\} \right\}   \right] \right\} \nonumber\\
&-\rho \sum_{j=1}^{p} \sum_{i=1}^{n} \{\bB(T_i)^\top\bgam_j\}^2,
\end{align}
forming the basis for estimation and inference.

Let $\tbgam = \argmax{\bgam} {\rm E}_{T,\Delta,\bZ} {\rm PL}(\bgam)$, where the expectation is taken with respect to the joint distribution of $T,\Delta$ and $\bZ$ under the true parameter $\bbeta_0(t)$. An estimate of $\tbgam$ is obtained by maximizing the likelihood \eqref{ch3:eq:PL3} so that
$\hbgam = \argmax{\bgam}\,{\rm PL}(\bgam).$
Then an estimate of $\bbeta(t)$ is given by $\hbbeta(t) = \{\hbeta_1(t), \ldots, \hbeta_{p}(t)\}^\top$ with $\hbeta_j(t) = \zeta{(\bB(t)^\top\hat{\bgam}_j,\alpha_j)}$.

Optimizing $\mathrm{PL}(\bgam)$ can be implemented by gradient-based methods
\citep{chau2014simulation}
and  coordinate descent algorithms \citep{wright2015coordinate}. With appropriate initial values, global optimizers can be reached. Specifically, for each $j=1,\ldots,p$, we obtain the non-varying coefficients $(a_1,\ldots,a_p)^\top$ from the Cox model, then we set the initial $\bgam^{(0)}_j$ to be a vector of $a_j$ with length $q$.  \yl{In practice, we recommend to vary the initial values and check the robustness of the final results.} We  choose the pre-specified parameters as follows. 
In theory, our method works for any ${\alpha}$; however, in practice,  a value  of $ {\alpha}$ comparable to the scale of true coefficients works best. Thus, we set $\alpha_j$ to be $ 0.5 \times |a_j|$. The choices of $\eta$ and $\rho$ can be specified in accordance with Condition \ref{ch3:con:rho}. The knots of B-spline are equally spaced over $[0,\tau]$. The number of basis functions, $q$,  can be determined through  
$R$-fold    cross-validation. That is, partition the full data $D$  into $R$ equal-sized groups, denoted by $D_r$, for $r = 1 \ldots, R$, and let $\hbbeta_{-r}^{(q)}(t)$ be the estimate obtained with $q$ bases  using all the data except for $D_r$. We obtain the optimal $q$  by minimizing the cross-validation error, which is the average of  the negative objective function
(\ref{ch3:eq:PL0})
evaluated at $\hbbeta_{-r}^{(q)}(t)$ on $D_r$ with $r$ running from   1 to $R$.
\section{Inference}\label{ch3:sec:asymp}
We begin with some needed notation.
\yl{First, for a $p_1 \times q_1$ matrix $A=(a_{ij})$ and a  $p_2 \times q_2$ matrix $B=(b_{ij})$, their Kronecker product is defined to be
\[
A \otimes B =\left( 
\begin{array}{ccc} 
a_{11} B & \ldots & a_{1q_1} B \\
\ldots & \ldots &  \ldots \\
a_{p_11} B & \ldots &  a_{p_1q_1} B
\end{array}
\right).
\]}
With that, we define the following:
\begin{align*} 
	\begin{split}
		g(\bbeta,\bZ,t) &= \sum_{j=1}^{p} Z_j\beta_j(t), \\ 
		g_n(\bgam,\bZ,t) & = \sum_{j=1}^{p} Z_j h_j(\bB(t)\bgam_{j}),\\
		S_{0n}(\tbgam,t) & = \frac{1}{n}  \sum_{i=1}^{n} Y_i(t) \exp(g_n(\tbgam,\bZ_i,t)), \\
		S_{0}(t) & = \rm E  Y(t) \exp(g(\bbeta,\bZ,t)), \\
		S_{1n}(\tbgam,t) & = \frac{1}{n}  \sum_{i=1}^{n} Y_i(t) \exp(g_n(\tbgam,\bZ_i,t)) \bZ_i\otimes\bB_i, \\
		S_{1}(t) & = \rm E  Y(t) \exp(g(\bbeta,\bZ,t))\bZ\otimes\bB, \\
		S_{2n}(\tbgam,t) & = \frac{1}{n}  \sum_{i=1}^{n} Y_i(t) \exp(g_n(\tbgam,\bZ_i,t)) (\bZ_i\bZ_i^\top)\otimes(\bB_i\bB_i^\top), \\
		\text{and } \quad S_{2}(t) & = \rm E  Y(t) \exp(g(\bbeta,\bZ,t))(\bZ\bZ^\top)\otimes(\bB\bB^\top),
	\end{split}
\end{align*}
followed by some key sufficient conditions that guarantee the properties of our estimator.
\begin{enumerate}[C1]
	\item \label{ch3:con:time} The failure time $T^u$ and the censoring time $T^c$ are conditionally independent given the covariate $\bZ$.
	\item \label{ch3:con:tau}  $\tau$ is chosen so that $\Pr( T^u > \tau | \bZ ) >0$ almost surely and $\tau <\infty$;
	at $\tau$, the baseline cumulative hazard function $\Lambda_0(\tau) \equiv \int_{0}^{\tau} \lambda_0(s)ds < \infty$.
	\item \label{ch3:con:z} The covariates $\bZ$ takes value in a bounded subset of $\mR^p$ and  $\Pr (Z_{j}=0) < 1$. Also,  $\sum_{j=1}^{p}|Z_j|=O_p(1)$.
	\item \label{ch3:con:index}There exists a small positive constant $\epsilon$ such that $\Pr(\Delta =1|\bZ)>\epsilon$ and $\Pr(T^c>\tau|\bZ) > \epsilon$ almost surely. 
	\item \label{ch3:con:density}Let $0<c_1<c_2< \infty$ be two constants. The joint density $f(t,\bz,\Delta=1)$ of $(T,\bZ,\Delta=1)$ satisfies $c_1 \leq f(t,\bz,\Delta=1) < c_2$ for all $(t,\bz) \in[0,\tau]\times \mR^{p}$.
	\item \label{ch3:con:rho} $\eta=o(q^{-m})$, $\rho=O(n^{a})$ with $a\le-1$, and $q=o(n)$.
	\item \label{ch3:con:s012}There exist a neighborhood $\Theta$ of $\tbgam$ and scalar, vector and matrix functions $s_{0}$, $s_{1}$ and $s_{2}$ defined on $\bgam \times [0,\tau]$ such that for $j=0,1,2$,
	\begin{align*}
		\sup_{0\leq t \leq\tau, \bgam \in \Theta} || S_{j}(\bgam,t)- s_{(j)}(\bgam,t)|| \rightarrow_p 0.
	\end{align*}
	\item \label{ch3:con:Sigma}Let $\Theta$, $s_{0}$, $s_{1}$ and $s_{2}$ be as in Condition \ref{ch3:con:s012} and define $e={s_{1}}/{s_{0}}$ and $v= {s_{2}}/{s_{0}}- e^{ \otimes 2}$. For all $\bgam \in \Theta$, $ t \in [0,\tau]$:
	\begin{align*}
		s_{1}(\bgam,t) = \frac{\partial}{\partial \bgam} s_{0} (\bgam,t), \quad 
		s_{2}(\bgam,t) = \frac{\partial^2}{\partial \bgam \partial  \bgam^\top} s_{0} (\bgam,t),
	\end{align*}
	$s_{0}(\cdot,t)$, $s_{1}(\cdot,t)$, $s_{2}(\cdot,t)$ are continuous functions of $\bgam \in \Theta$, uniformly in $t \in [0,\tau]$, $s_{0}$, $s_{1}$, and $s_{2}$ are bounded on $\Theta \times [0,\tau]$, and the matrix 
	\begin{align*}
		\Sigma(\tbgam,\tau) = \int_{0}^{\tau} v(\tbgam,t) s_{0}(\tbgam,t) \tbgam(t) dt
	\end{align*}
	is positive definite.
	\item \label{ch3:con:Linde}There exists a $\delta>0$ such that
	$$ n^{-1/2} \sup_{i,t}|| \bZ_i||_{\infty}|Y_i(t)\mI\{\bZ_i^\top\bbeta>-\delta|| \bZ_i||_{\infty} \} | \rightarrow_p 0.$$
\end{enumerate}
Condition \ref{ch3:con:time} is commonly assumed in survival analysis  for non-informative censoring. The finite $\tau$ condition  of \ref{ch3:con:tau} is assumed in many studies, including \cite{Anderson1982}. Condition \ref{ch3:con:z} is often assumed in nonparametric regression and is reasonable in practical situations as we do not observe infinite covariates. Condition \ref{ch3:con:index} controls the censoring rate so that the data have adequate information \cite{sasieni1992non}. Condition \ref{ch3:con:density} is needed for model identifiability and used in Huang (1999) \cite{Huang1999}. Condition \ref{ch3:con:rho} controls  estimation biases and ensures the convergence. Conditions \ref{ch3:con:s012}, \ref{ch3:con:Sigma}, and \ref{ch3:con:Linde} are regularity conditions, which can be found in Anderson and Gill (1982) \cite{Anderson1982}.

\subsection{Asymptotic theory}

\begin{theorem}\label{ch3:thm:converg}
	Suppose Conditions \ref{ch3:con:time}-\ref{ch3:con:rho} hold. If $\bbb_{0j}(t)\in \mS_{q,\alpha_j}$  for $j=1,\ldots,p$ with $q$ and $\alpha_j$ being the same as in $ {\rm PL}(\btheta)$,  then
	$$||\hbbeta- \bbeta_0||_2= O_p\left((q/n)^{1/2}\right);$$
	if $\bbb_{0j}(t)\notin \mS_{q,\alpha_j}$ for $j=1,\ldots,p$,
	$$||\hbbeta- \bbeta_0||_2= O_p\left(r_n^{1/2}\right),$$
	where $r_n = q/n+q^{-2m}$.
\end{theorem}

Theorem \ref{ch3:thm:converg}  implies convergence  of $\hbbeta$ by Condition \ref{ch3:con:rho} and $m>\mbox{0$\cdot$5}$.  If the true curves are in the thresholding sieve space,  there is no approximation error; and if $q$ is  $O(1)$, Theorem \ref{ch3:thm:converg} suggests root-$n$ consistency.

Let $\be_j$ be a directional vector of length $p$ with $j$th entry as 1 and others 0. For any $t\in[0,\tau]$, let $\ba(t)= \be_j\otimes\bB(t)$, then $\htheta_j(t)  =\ba(t)^\top\hat{\bgam}$. 
\begin{theorem}\label{ch3:thm_asymp}
	Under Conditions  \ref{ch3:con:time}-\ref{ch3:con:Linde}, we have for any $t\in [0,\tau]$ and $j = 1, \ldots,p$, 
	\begin{align*}
		\frac{\hat{\theta}_j(t)-\theta_j(t)}{\sigma_{nj}(t)} \rightarrow_d N(0,1), \quad \text{as } n \rightarrow \infty,
	\end{align*}
	where $\sigl^2(t) =  n\ba(t)^\top \left[\{\rm PL\}^{''} (\tbgam) \right]^{-1}\Sigma(\tbgam,1)\left[\{\rm PL\}^{''} (\tbgam) \right]^{-1}\ba(t)$.
\end{theorem}

With Theorem \ref{ch3:thm_asymp}, we can then obtain the asymptotic distribution of $\hbeta_j(t)$ based on $\hbeta_j(t)= \zeta{\{\htheta_j(t),\alpha_j\}}$.
\begin{theorem}\label{ch3:thm:cdf}
	Under Conditions  \ref{ch3:con:time}--\ref{ch3:con:Linde}, for  any $t \in[0,\tau]$, the limiting distribution of $\hbeta_j(t)$ {\rm(} $j = 1, \ldots, p${\rm)} satisfies 
	$$\lim_{n\rightarrow \infty}\Bigg| \mathrm{Pr}(\hbeta_j(t)\leq x)  - G_{nj}(x)\Bigg|  \ =\ 0,
	$$ 
	where $G_{nj}(x)=\left[\Phi\left\{\frac{x+\alpha_j-\ttheta_j(t)}{\sigl(t)}\right\}\mI(x \geq 0) +
	\Phi\left\{\frac{x-\alpha_j-\ttheta_j(t)}{\sigl(t)}\right\} \mI(x<0)\right] $ and $\Phi(\cdot)$ is the cumulative distribution function of $N(0,1)$.
\end{theorem}

The limiting  distribution in Theorem \ref{ch3:thm:cdf} guarantees that our proposed estimator can detect the zero-effect regions, because the probability of $\hbeta_j(t)=0$ can be greater than 0 even with a finite sample size. 

\subsection{Sparse confidence intervals}
We introduce the sparse confidence intervals to gauge the uncertainty of the point estimates and make valid statistical inferences on the selection and the zero-effect region detection. 

Given a $\xi \in (0,1)$, for any $t\in [0,\tau]$ we construct a pointwise $(1-\xi)$ level asymptotic sparse confidence interval  for $\bbb_j(t)$, denoted by $[u_{nj}(t),v_{nj}(t)]$.  Let $z_{\xi/2}$ and $\Phi$ be the $(1-\xi/2)$ quantile and the cumulative distribution function of $N(0,1)$, respectively.  Let $P_{+} = \Pr\{\hbeta_j(t) > 0\}$ and $P_{-} = \Pr\{\hbeta_j(t)< 0\}$, which can be estimated by  $\hat{P}_{+} = 1-\Phi\{(\alpha_j-\htheta_j)/\hsigl(t)\}$ and $\hat{P}_{-} = \Phi\{(-\alpha_j-\htheta_j)/\hsigl(t)\}$ using Theorem \ref{ch3:thm:cdf}. Here,
$\hsigl(t)$ is as defined in Theorem \ref{ch3:thm_asymp}. We   construct $[u_{nj}(t),v_{nj}(t)]$ as follows:
	
	\begin{itemize}
		\item  if $\hat P_{+}  + \hat P_{-}   \le  \xi $, $u_{nj}(t) = v_{nj}(t) = 0$;
		\item else if  $\hat P_{+} < \xi/2$ and $\hat P_{-} <1-\xi/2 $, $[u_{nj}(t),v_{nj}(t)] = \big[\hbeta_j(t)- \hsigl(t) \hat B ,  0\big] $ with $\hat B= \Phi^{-1}\big\{1-\xi+\Phi(-\hsigl^{-1}(t)\alpha_j+\hsigl^{-1}(t)\htheta_j)\big\}$;
		\item else if  $\hat P_{-} < \xi/2$ and $\hat P_{+} <1-\xi/2 $,  $[u_{nj}(t),v_{nj}(t)] = \big[0, \hbeta_j(t)+\hsigl(t) \hat A\big]$ with $\hat A= -\Phi^{-1}\big\{\xi-1+\Phi(\hsigl^{-1}(t)\alpha_j+\hsigl^{-1}(t)\htheta_j)\big\}$;
		\item else  $[u_{nj}(t),v_{nj}(t)] = \big[\hbeta_j(t)- \hsigl(t) z_{\xi/2}  , \hbeta_j(t)+ \hsigl(t) z_{\xi/2}\big]$.
	\end{itemize}

	\begin{theorem}\label{ch3:thm:SCI}
		Under Conditions \ref{ch3:con:time}-\ref{ch3:con:Linde}, $[u_{nj}(t), v_{nj}(t)]$ is a $(1-\xi)$ level sparse confidence interval of $\beta_j(t)$ for $j=1,\ldots,p$ and any $t \in[0,\tau]$. 
	\end{theorem}
	
	We omit its proof as it is a straightforward application of Theorem \ref{ch3:thm:cdf}.

	\section{Simulations}\label{ch3:sec:simul}
	We  compare the proposed model with the regular time-varying effects Cox model. With $p=3$, we design some special varying coefficient functions  containing zero-effect regions as follows:
	\begin{eqnarray}\label{ch3:timevaryfun}
	& & \beta_1(t)= (-t^2+3)\mI(t\leq \sqrt{3}),                \nonumber    \\
	& &\beta_2(t)=2\log(t+0.01)\mI(t\ge 1),                \\
	\text{and }\quad & &\beta_3(t)=\left(\frac{-6}{t+1}+2\right)\mI(t\le 2). \nonumber 
	\end{eqnarray}
	
	We first simulate $\bm{Z_i}=(Z_{i1},Z_{i2},Z_{i3})^\top\sim {\rm N}(\bm{0},\Sigma)$, where $\Sigma$ has the following three structures: independent (Ind) with $\mathrm{cov}( Z_{ij},  Z_{ij^*}) = \mI(j=j^*)$, autoregressive [AR(1)] with $\mathrm{ cov}( Z_{ij},  Z_{ij^*}) = \mbox{0$\cdot$5}^{|j-j^*|}$, and  compound symmetry  (CS) with $\mathrm{ cov}( Z_{ij},  Z_{ij^*}) = \mI(j=j^*)+\mbox{0$\cdot$5}   \mI(j\neq j^*)$. We simulate $U_i \sim U(0,1)$, and solve $T^u_i$ using $U_i=1-\exp\left\{-\int_{0}^{T_i^u}\lambda_0(u) \exp({\sum_{j=1}^{3}Z_j\beta_j(u)}){\rm d}u\right\}$, where $\lambda_0(u)$ is set to be some constant in $(0,1)$. The censoring times $C_i$ are generated from $U(0,10)$, and $T_i^c= C_i \wedge 3$. 
	
	We choose sample sizes $n=500$, $2,000$ and $5,000$, and generate 200 independent datasets for each setting. For implementing our proposed model, we set $d=3$, $\eta = \mbox{0$\cdot$001}$, and $\alpha_j$ to be half of the absolute values of the least-squares estimates. The number of knots, $K$, is selected via cross-validation. \yl{Specifically, we tune $K$ by conducting 10-fold cross-validation over a reasonable range such as \{3, 5, 9, 13, 17, 21\}. We  note that $\rho$ can also be selected via cross-validation;  however, that would increase the computational burden, given that  $K$  needs to to be tuned. Based on our
numerical experience, we have found that specifying $\rho=1/n^2$ that satisfies Condition C6 would give a good performance. Therefore, we set $\rho=1/n^2$ in simulations and our later data analysis.}
	
	For comparison, we  fit   time-varying effects Cox models by following \cite{Zucker1990, Yan2012}.  For evaluation criteria, we use the integrated squared errors (ISE) and the averaged integrated squared errors (AISE), defined as ${\rm ISE}(\bbb_j) = n_g^{-1} \sum_{g = 1}^{n_g} \{\hbeta_j(t_g) - \bbb_j(t_g)\}^2$ and ${\rm AISE} = p^{-1}\sum_{j=1}^{p} {\rm ISE}(\bbb_j)$, respectively, where $t_g$ ($g=1,\ldots,n_g$) are the grid points on $(0,3)$. Table \ref{ch3:tab:mse}  shows that the soft-thresholding time-varying effects Cox model has a better accuracy than the regular time-varying effects  Cox model by presenting
	smaller integrated squared errors and averaged integrated squared errors.
	\begin{table}[!htb]
		\centering
		\begin{threeparttable}
			\caption{Comparisons of estimation accuracy for the soft-thresholding time-varying effects  Cox model and  the regular time-varying effects  Cox model.} \label{ch3:tab:mse}
			\small
			\begin{tabular}{rllllll}
				\hline
				Covariance & n& Model & ISE$(\beta_1)$ & ISE$(\beta_2)$ & ISE$(\beta_3)$ & AISE \\ 
				\hline
				&&&&&&\\[-0.5ex]
				& 500 & STTV & 62.6 (77.1) & 53.1 (43.5) & 58.7 (59.5) & 58.1 (39.7) \\ 
				&  & RegTV & 75.5 (94.6) & 56.6 (44.3) & 61.9 (60.6) & 65.4 (46.2) \\ 
				&&&&&&\\[-2ex]
				Ind & 2000 & STTV & 12.4 ( 9.7) & 12.0 ( 8.5) & 13.1 (10.4) & 12.5 ( 5.7) \\ 
				& & RegTV & 13.9 ( 8.2) & 11.8 ( 8.6) & 12.4 ( 8.8) & 12.7 ( 5.1) \\ 
				&&&&&&\\[-2ex]
				& 5000 & STTV & 4.2 (3.2) & 4.1 (2.8) & 4.0 (2.7) & 4.1 (1.7) \\ 
				&  & RegTV & 5.6 (3.0) & 4.2 (2.8) & 4.5 (2.7) & 4.7 (1.6) \\ 
				&&&&&&\\[-0.5ex]
				& 500 & STTV & 16.2 (16.0) & 18.2 (47.1) & 15.2 (11.1) & 16.5 (18.7) \\ 
				&  & RegTV & 16.3 (14.1) & 20.9 (50.3) & 13.4 ( 8.2) & 16.9 (19.4) \\ 
				&&&&&&\\[-2ex]
				AR(1) & 2000 & STTV & 3.6 (2.2) & 2.6 (2.0) & 3.9 (2.3) & 3.3 (1.5) \\ 
				&  & RegTV & 3.7 (2.2) & 2.8 (2.5) & 3.1 (1.6) & 3.2 (1.4) \\
				&&&&&&\\[-2ex] 
				& 5000 & STTV & 1.3 (1.0) & 1.1 (0.9) & 1.2 (0.8) & 1.2 (0.6) \\ 
				& & RegTV & 1.9 (0.9) & 1.3 (0.9) & 1.3 (0.8) & 1.5 (0.6) \\ 
				&&&&&&\\[-0.5ex]
				& 500 & STTV & 18.9 (24.6) & 19.1 (30.2) & 16.5 (14.6) & 18.2 (16.2) \\ 
				&   & RegTV & 19.1 (15.5) & 20.4 (30.3) & 17.0 (12.2) & 18.8 (13.2) \\ 
				&&&&&&\\[-2ex]
				CS & 2000 & STTV & 3.6 (2.6) & 2.7 (2.5) & 3.8 (2.7) & 3.4 (1.8) \\ 
				&   & RegTV & 4.0 (2.3) & 2.8 (2.4) & 3.2 (1.6) & 3.4 (1.4) \\ 
				&&&&&&\\[-2ex]
				& 5000 & STTV & 1.2 (0.8) & 1.1 (0.9) & 1.0 (0.6) & 1.1 (0.5) \\ 
				&  & RegTV & 1.8 (0.7) & 1.1 (0.9) & 1.2 (0.7) & 1.4 (0.5) \\ 
				\hline
			\end{tabular}
			\begin{tablenotes}
				\small
				\item   STTV: the soft-thresholding time-varying effects  Cox model; RegTV: the regular time-varying effects  Cox model; ISE: the integrated squared errors; AISE: the averaged integrated squared errors. All numbers are after being multiplied by 100.
			\end{tablenotes}
		\end{threeparttable}
	\end{table}
	
\begin{figure}[!htbp]
		\centering
		\small
		\includegraphics[width=0.45\textwidth]{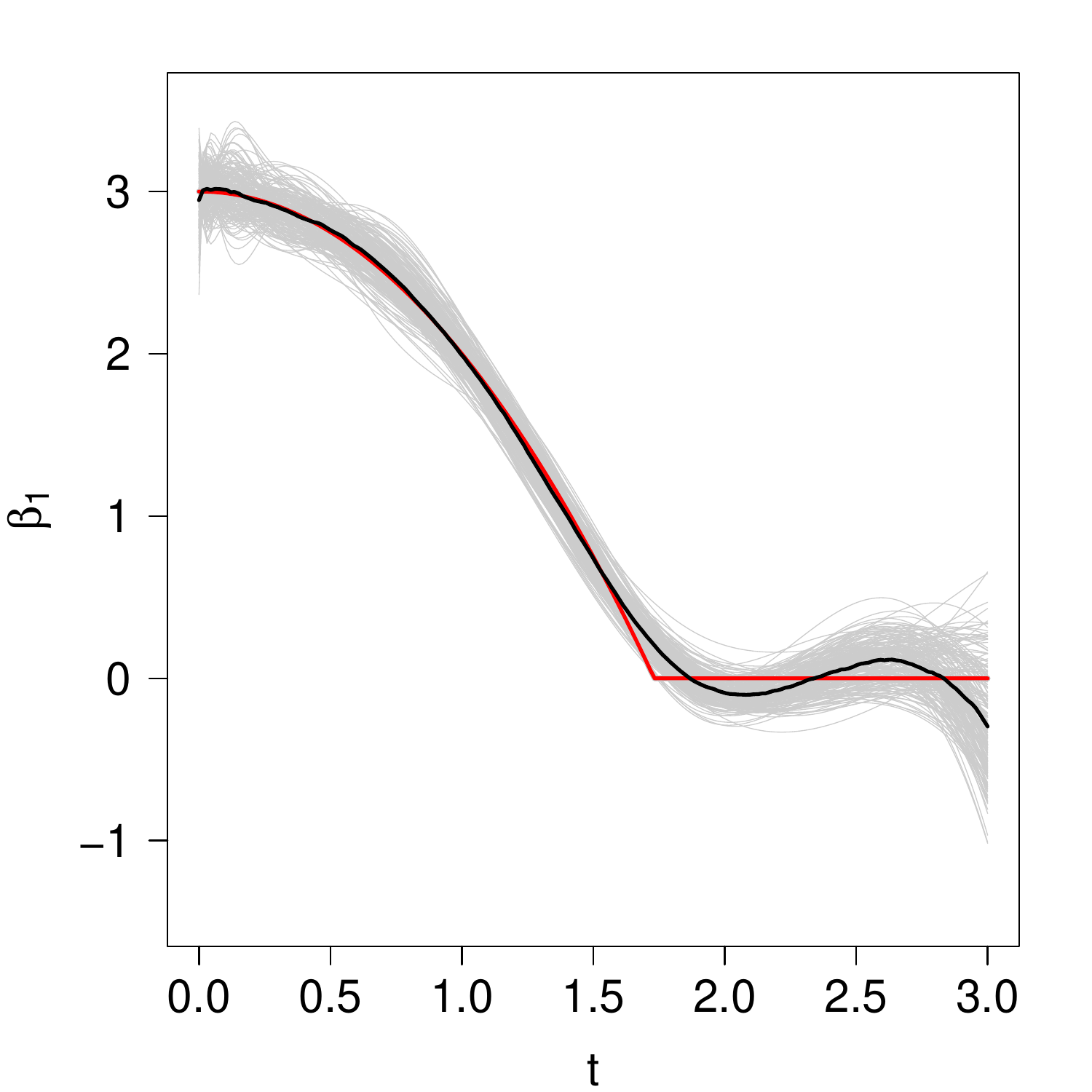}
		\includegraphics[width=0.45\textwidth]{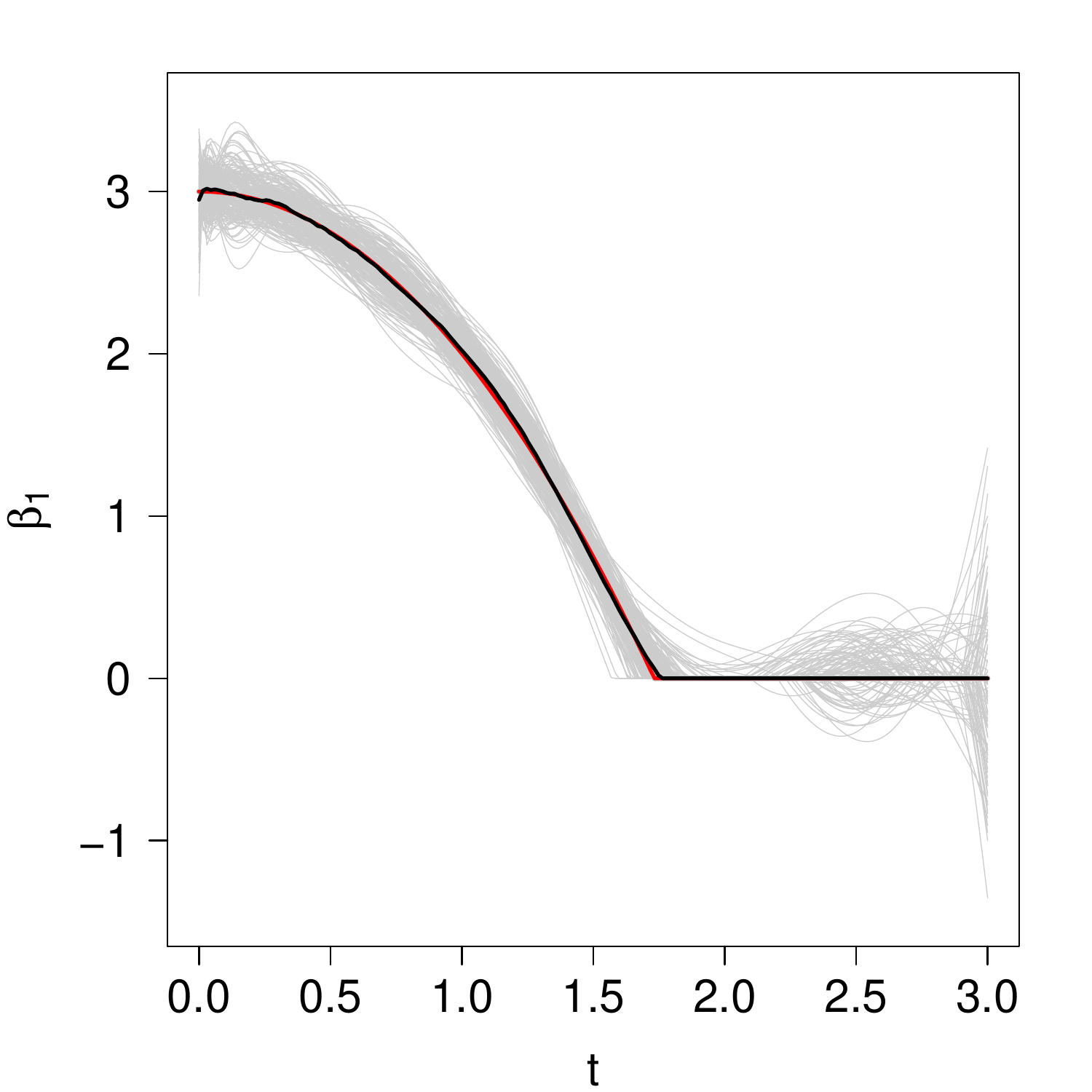}\\
		\includegraphics[width=0.45\textwidth]{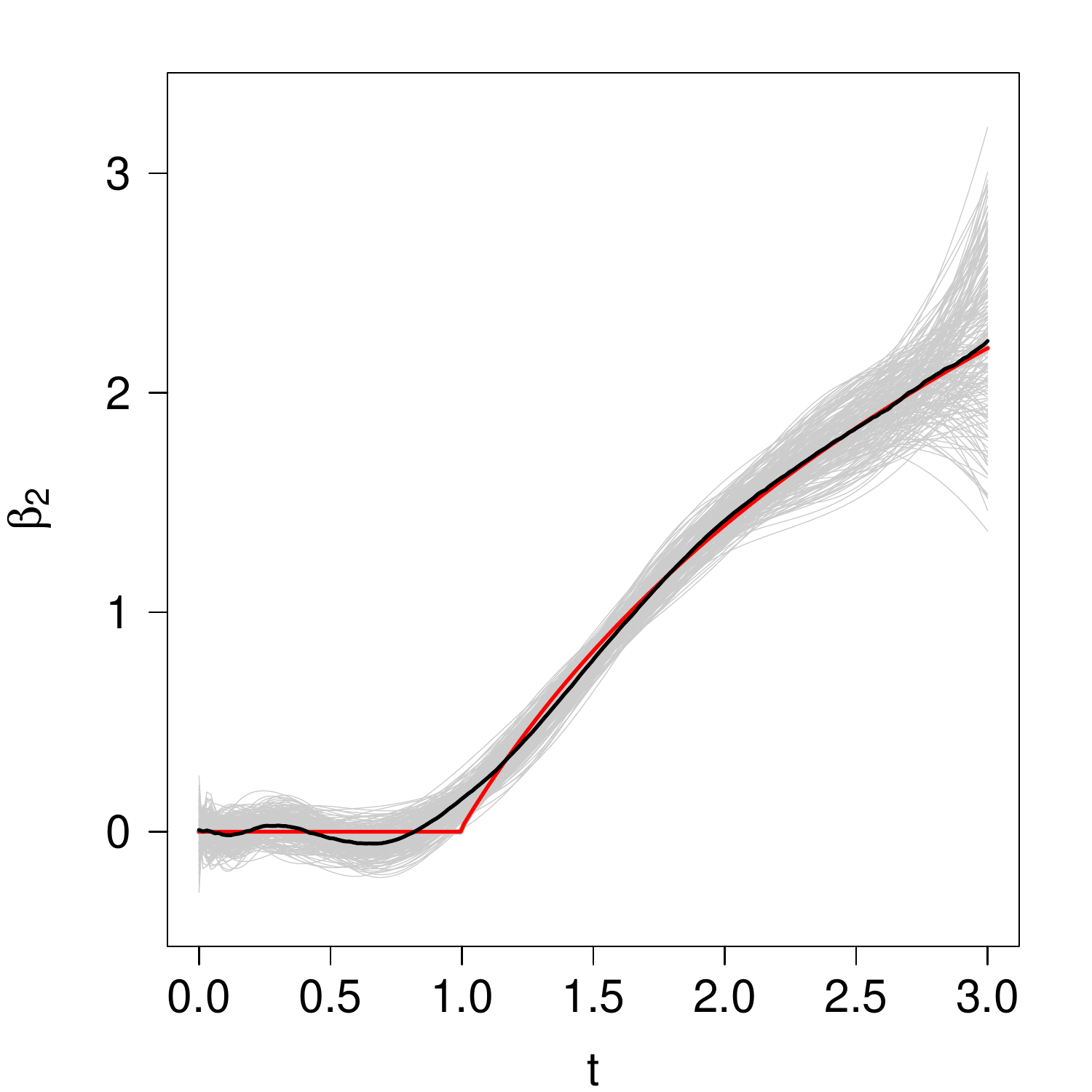}
		\includegraphics[width=0.45\textwidth]{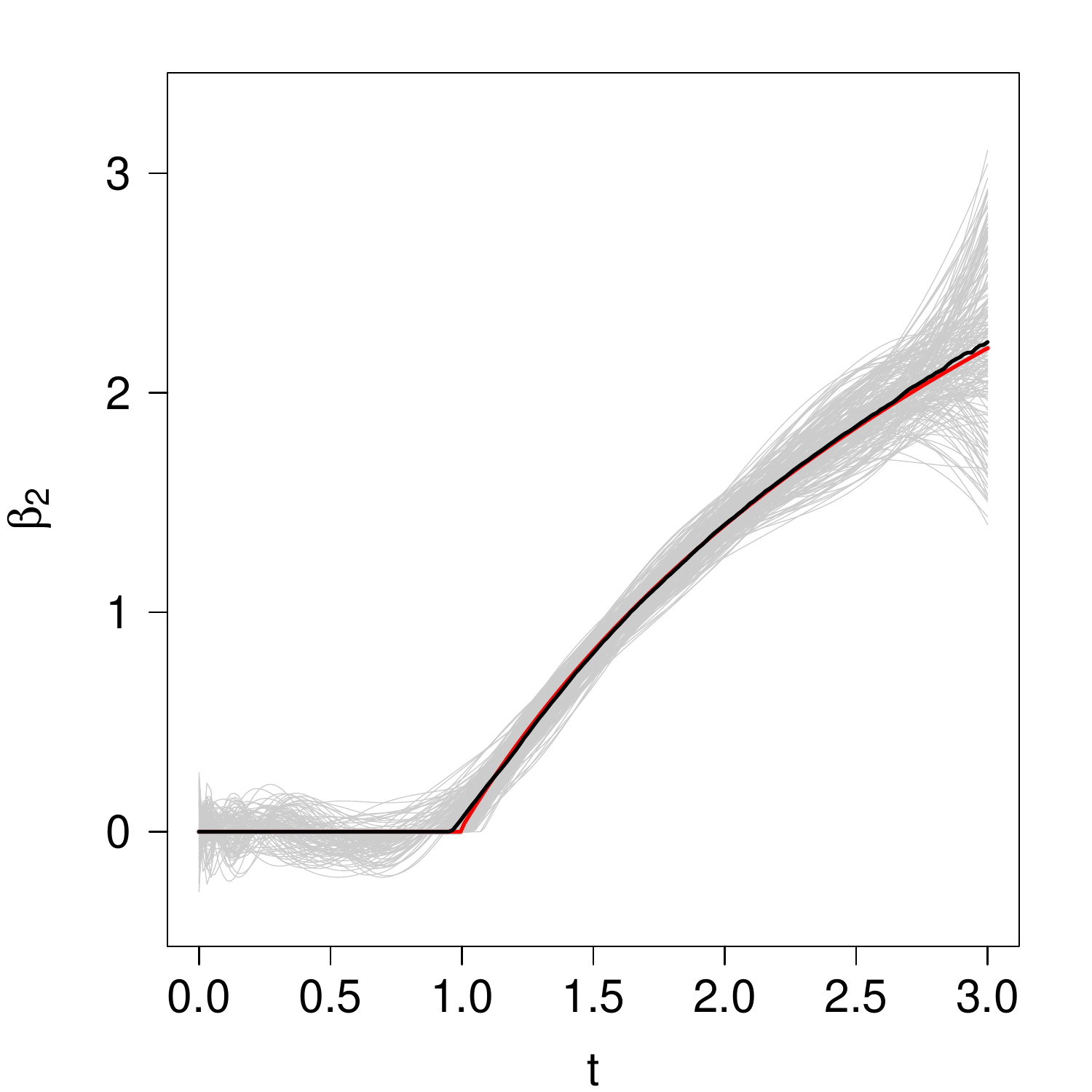}\\
		\includegraphics[width=0.45\textwidth]{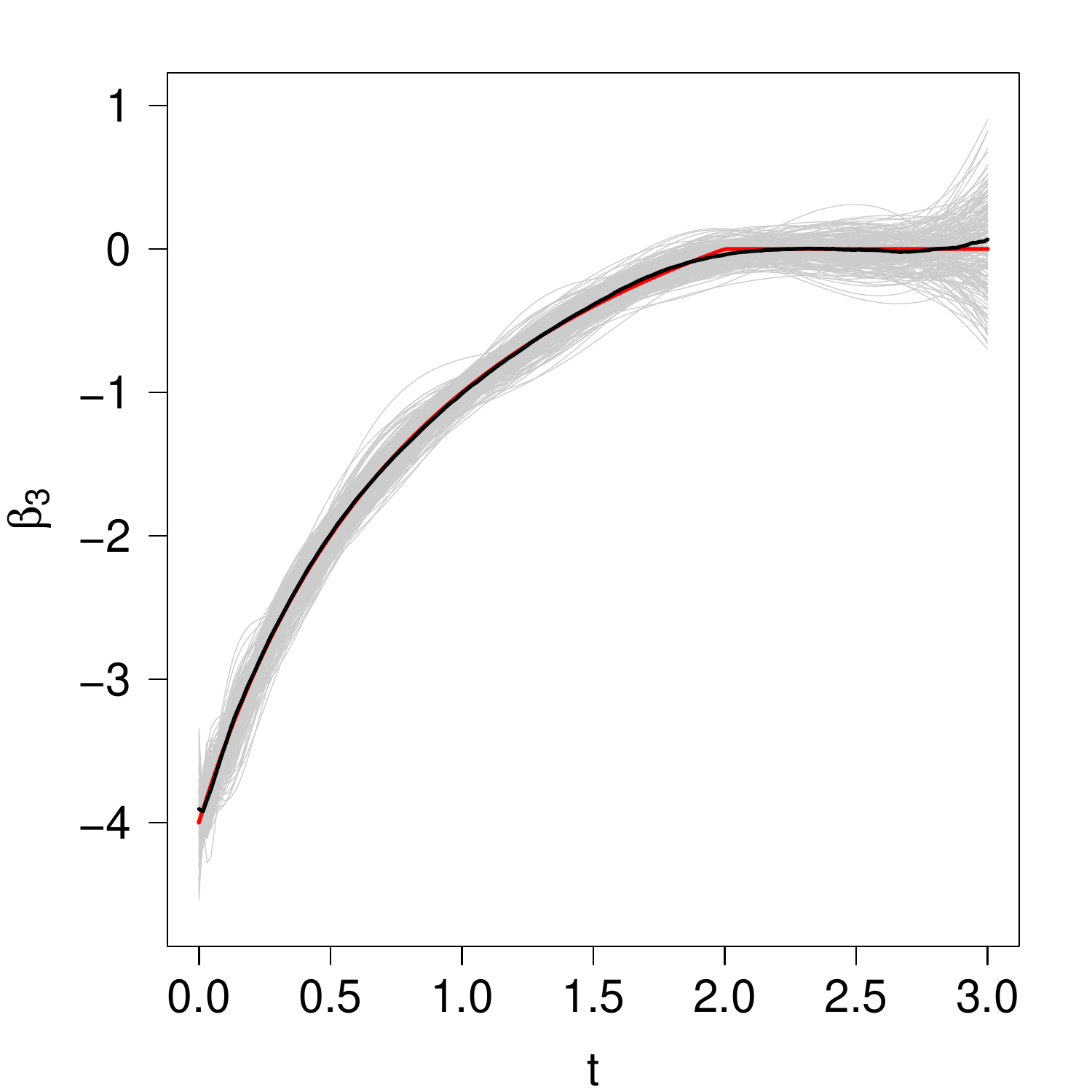}
		\includegraphics[width=0.45\textwidth]{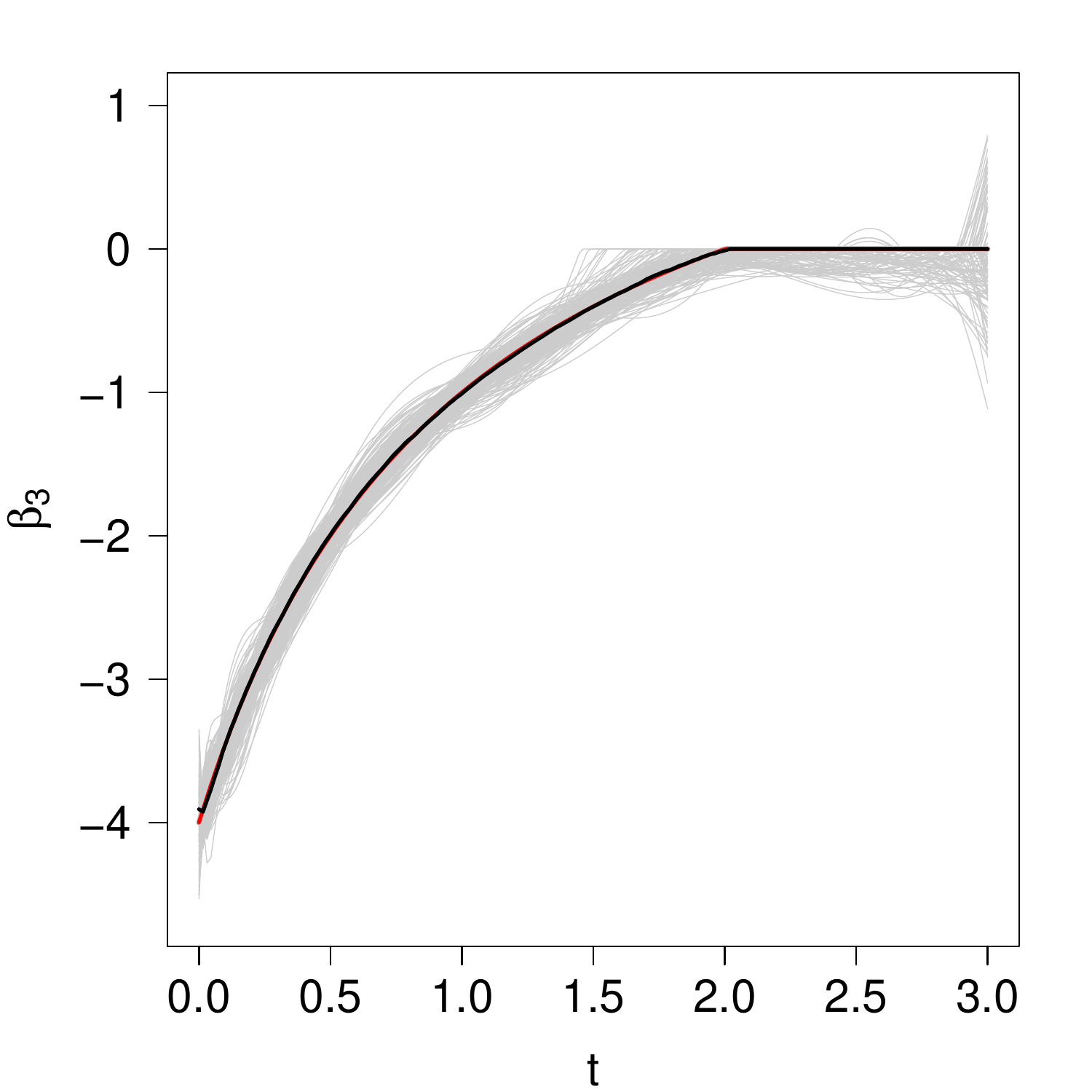}\\
		\caption{Comparisons of the results obtained from the soft-thresholding time-varying effects  Cox model (right panel) and the regular time-varying effects  Cox model (left panel); the gray curves are 200 estimated curves based on 200 simulations, the black curves are the median estimates, and the red curves are the  truth; the sample size is $5,000$ and the average censoring rate is 0.12.}    \label{ch3:fig:md}
	\end{figure}

	Figure \ref{ch3:fig:md}, which plots the estimation curves and their median for the soft-thresholding time-varying effects  Cox model and the regular time-varying effects  Cox model, shows the medium estimation curves from the soft-thresholding time-varying effects  Cox model coincide with the truth and the soft-thresholding approach has the zero-effect detection ability.
In contrast,  the regular time-varying effects  Cox model fails to estimate zero effects.  
	
\begin{figure}[!htbp]
		\centering
		\small
		\includegraphics[width=0.32\textwidth]{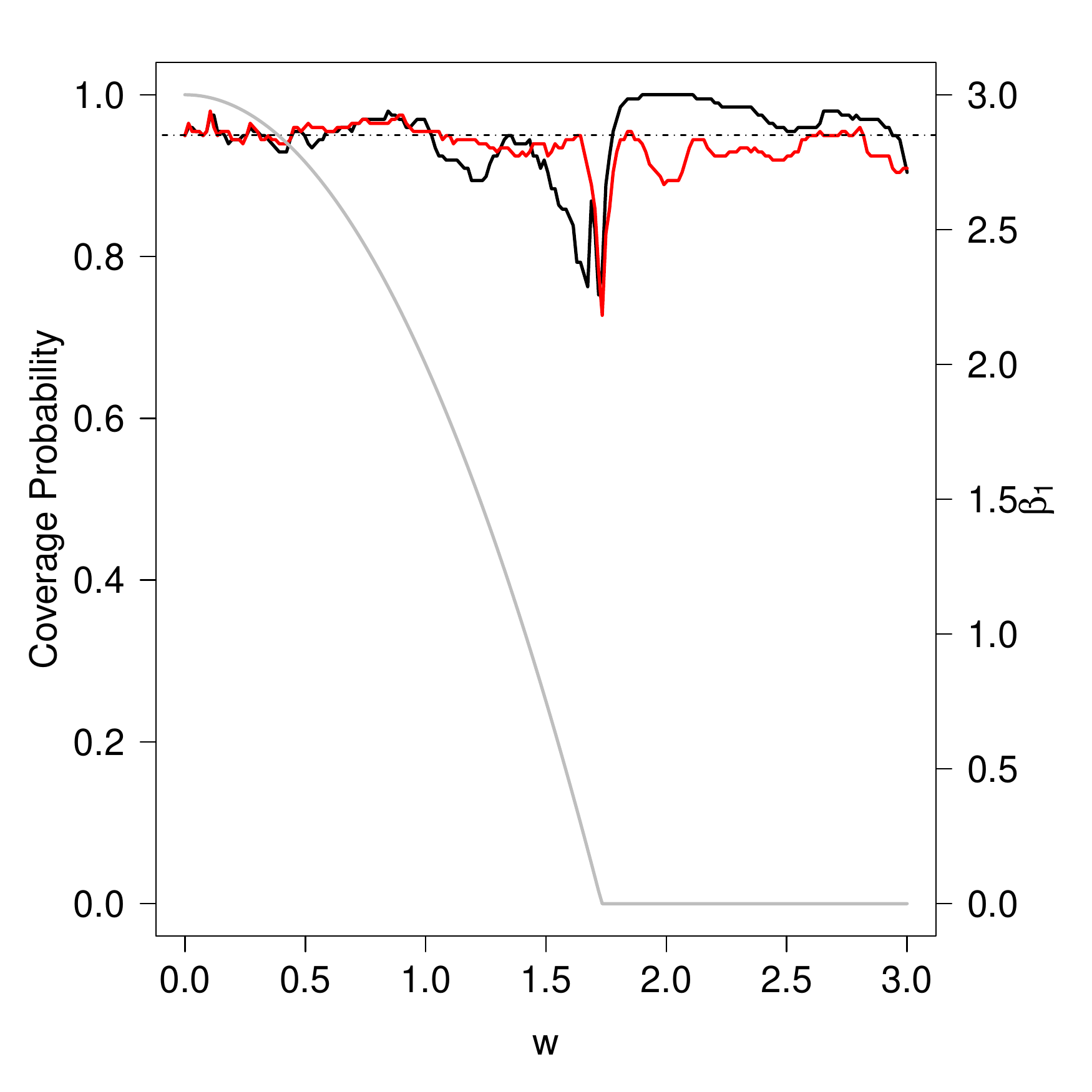}
		\includegraphics[width=0.32\textwidth]{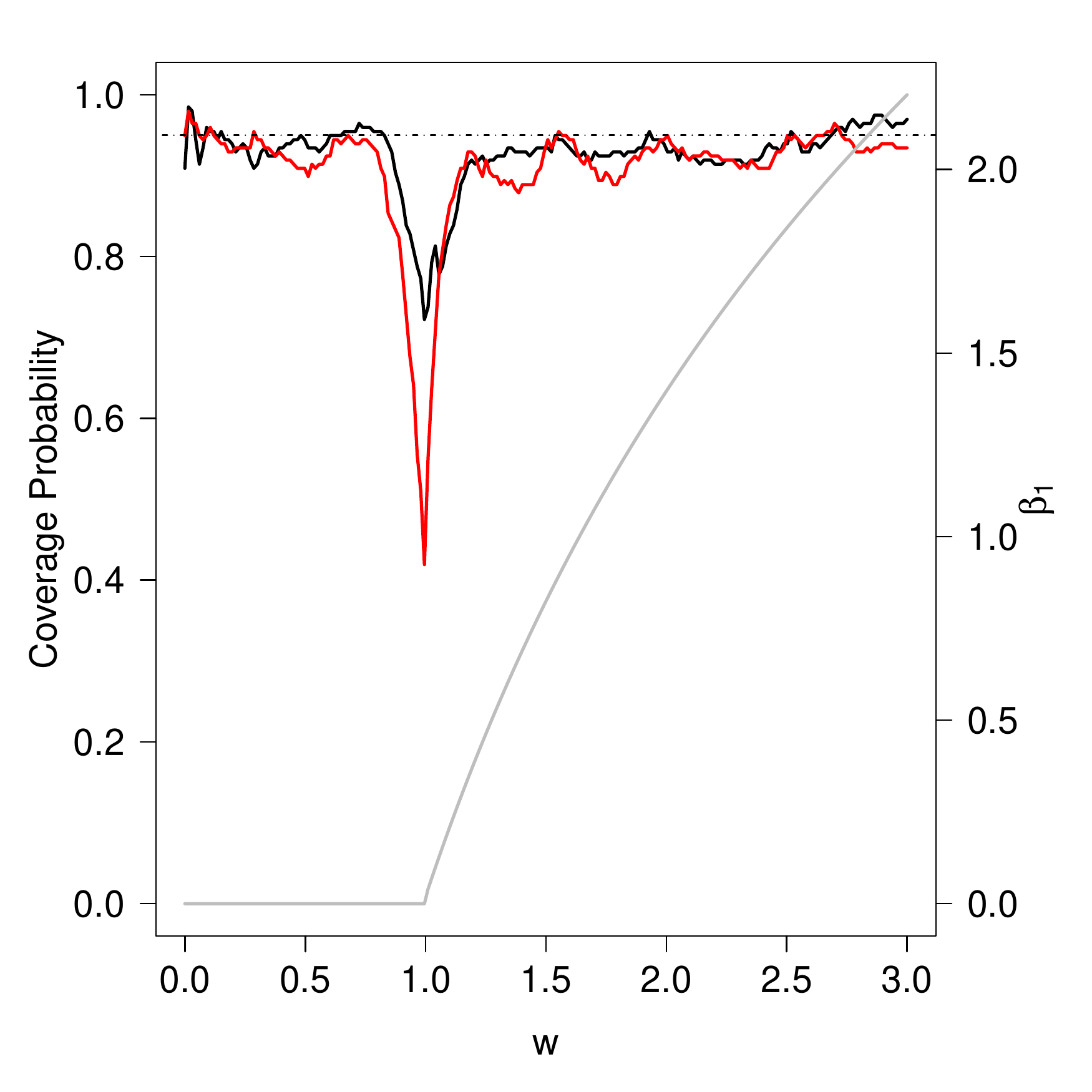}
		\includegraphics[width=0.32\textwidth]{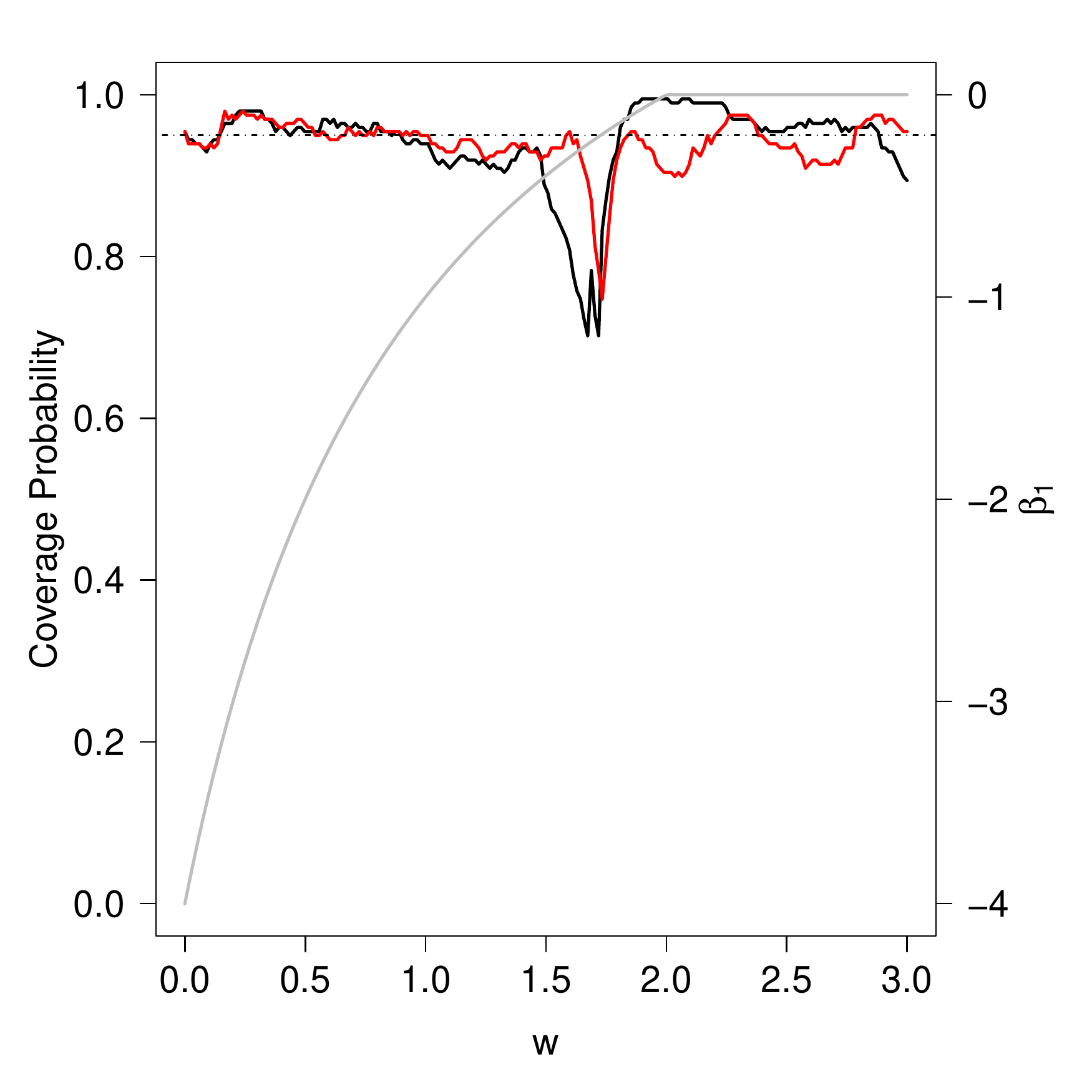}
		\caption{Comparisons of coverage probability from the regular time-varying effects  Cox model (RegTV) \yl{(the black curve)} and the soft-thresholding time-varying effects  Cox model (STTV) \yl{(the red curve)}. The  sample size is $5, 000$ and the average censoring rate is 0.12.}    \label{ch3:fig:cp}
	\end{figure}

	Figure \ref{ch3:fig:cp} compares the estimated coverage probabilities from the soft-thresholding time-varying effects  Cox model and the regular time-varying effects  Cox model, and shows that the soft-thresholding time-varying  Cox model has a reasonable coverage probability in both zero-effect regions and non-zero-effect regions. In the region around the transition point, the soft thresholding time-varying effects  Cox model has a higher coverage probability estimation than the regular time-varying effects  Cox model. All of the results confirm that the soft-thresholding time-varying effects  Cox model draws better inference than the regular time-varying effects  Cox model.  
	
	With $|A|$ being the cardinality of set $A$, 
	we next compare zero-effect region detection using the following criteria:  
	\begin{align*}
		\begin{split}
			\text{Estimation-based true positive ratio: } {\rm ETPR}(\bbb) &=\frac{ |\{t:\hbeta(t)\ne0 \text{ and } \bbb(t) \ne 0\}| }{| \{t:\bbb(t) \ne 0\}|},\\
			\text{Estimation-based true negative ratio: }   {\rm ETNR}(\bbb) &=\frac{|\{t:\hbeta(t)=0\text{ and } \bbb(t) = 0\}|}{|\{t:\bbb(t)=0\}|}, \\
			\text{Inference-based true positive ratio: } {\rm ITPR}(\bbb) &=\frac{ |\{t: 0 \notin {\rm CI}\{\hbeta(t)\} \text{ and } \bbb(t) \ne 0\}| }{| \{t:\bbb(t) \ne 0\}|},\\
			\text{and} \qquad \qquad \qquad \qquad \qquad & \\
			\text{Inference-based true negative ratio: } {\rm ITNR}(\bbb) &=\frac{ |\{t: 0 \in {\rm CI}\{\hbeta(t)\}\text{ and } \bbb(t) = 0\}| }{|\{t:\bbb(t)=0\}|},
		\end{split}
	\end{align*}
	where ${\rm CI}\{\hbeta(t)\}$ is the 95\% confidence interval of $\bbb(t)$. 
	
	\begin{table}[!htb]
	\footnotesize
		\centering
		\begin{threeparttable}
			\caption{Comparisons of true positive ratios and true negative ratios for zero-effect region detection} \label{ch3:tab:TPRTNR}
			\begin{tabular}{lllcccccc}
				\hline
				&&&\multicolumn{4}{c}{STTV}&\multicolumn{2}{c}{RegTV}\\
				& n & $\beta$ & ETPR   & ETNR   & ITPR   & ITNR   & ITPR   & ITNR  \\ 
				\hline
				&&&&&&&&\\[-0.5ex]
				&   & $\beta_1$ & 0.96 (0.08) & 0.44 (0.25) & 0.81 (0.10) & 0.94 (0.11) & 0.81 (0.09) & 0.95 (0.11) \\ 
				& 500 & $\beta_2$ & 0.96 (0.05) & 0.22 (0.13) & 0.57 (0.17) & 0.94 (0.08) & 0.57 (0.18) & 0.94 (0.12) \\ 
				&   & $\beta_3$ & 0.95 (0.12) & 0.37 (0.28) & 0.55 (0.12) & 0.94 (0.12) & 0.55 (0.12) & 0.94 (0.12) \\ 
				&&&&&&&&\\[-2ex]
				&   & $\beta_1$ & 0.95 (0.06) & 0.61 (0.23) & 0.89 (0.07) & 0.95 (0.10) & 0.91 (0.06) & 0.94 (0.10) \\ 
				Ind & 2000 & $\beta_2$ & 0.97 (0.04) & 0.34 (0.16) & 0.85 (0.08) & 0.94 (0.08) & 0.86 (0.08) & 0.95 (0.08) \\ 
				&   & $\beta_3$ & 0.96 (0.12) & 0.50 (0.24) & 0.70 (0.10) & 0.94 (0.11) & 0.72 (0.10) & 0.95 (0.13) \\ 
				&&&&&&&&\\[-2ex]
				&   & $\beta_1$ & 0.98 (0.03) & 0.64 (0.27) & 0.95 (0.04) & 0.94 (0.10) & 0.96 (0.04) & 0.93 (0.11) \\ 
				& 5000 & $\beta_2$ & 0.98 (0.03) & 0.46 (0.18) & 0.92 (0.05) & 0.94 (0.09) & 0.93 (0.04) & 0.94 (0.10) \\ 
				&   & $\beta_3$ & 0.97 (0.09) & 0.50 (0.31) & 0.81 (0.09) & 0.96 (0.10) & 0.80 (0.08) & 0.96 (0.10) \\ 
				&&&&&&&&\\[-0.5ex]
				&   & $\beta_1$ & 0.96 (0.05) & 0.60 (0.22) & 0.90 (0.07) & 0.95 (0.10) & 0.93 (0.06) & 0.93 (0.12) \\ 
				& 500 & $\beta_2$ & 0.98 (0.04) & 0.32 (0.18) & 0.85 (0.08) & 0.92 (0.13) & 0.86 (0.08) & 0.93 (0.12) \\ 
				&   & $\beta_3$ & 0.97 (0.14) & 0.51 (0.27) & 0.69 (0.14) & 0.95 (0.13) & 0.73 (0.12) & 0.95 (0.12) \\ 
				&&&&&&&&\\[-2ex]
				&   & $\beta_1$ & 0.97 (0.04) & 0.71 (0.19) & 0.94 (0.04) & 0.95 (0.08) & 0.99 (0.02) & 0.92 (0.11) \\ 
				AR(1) & 2000 & $\beta_2$ & 0.99 (0.02) & 0.49 (0.19) & 0.95 (0.04) & 0.94 (0.10) & 0.97 (0.03) & 0.93 (0.11) \\ 
				&   & $\beta_3$ & 0.96 (0.09) & 0.62 (0.25) & 0.77 (0.10) & 0.92 (0.13) & 0.86 (0.07) & 0.94 (0.13) \\ 
				&&&&&&&&\\[-2ex]
				&   & $\beta_1$ & 1.00 (0.01) & 0.79 (0.17) & 0.98 (0.02) & 0.96 (0.08) & 1.00 (0.00) & 0.85 (0.11) \\ 
				& 5000 & $\beta_2$ & 1.00 (0.01) & 0.56 (0.17) & 0.98 (0.02) & 0.94 (0.09) & 1.00 (0.01) & 0.87 (0.11) \\ 
				&   & $\beta_3$ & 0.97 (0.05) & 0.63 (0.30) & 0.90 (0.05) & 0.97 (0.09) & 0.91 (0.05) & 0.96 (0.10) \\ 
				&&&&&&&&\\[-0.5ex]
				&   & $\beta_1$ & 0.96 (0.06) & 0.58 (0.23) & 0.90 (0.07) & 0.96 (0.10) & 0.92 (0.07) & 0.94 (0.12) \\ 
				& 500 & $\beta_2$ & 0.98 (0.03) & 0.32 (0.19) & 0.85 (0.07) & 0.93 (0.12) & 0.86 (0.07) & 0.94 (0.13) \\ 
				&   & $\beta_3$ & 0.98 (0.13) & 0.51 (0.29) & 0.70 (0.13) & 0.96 (0.11) & 0.71 (0.12) & 0.95 (0.11) \\ 
				&&&&&&&&\\[-2ex]
				&   & $\beta_1$ & 0.97 (0.04) & 0.68 (0.21) & 0.94 (0.04) & 0.96 (0.08) & 0.98 (0.02) & 0.92 (0.12) \\ 
				CS & 2000 & $\beta_2$ & 0.99 (0.02) & 0.48 (0.18) & 0.96 (0.04) & 0.94 (0.09) & 0.97 (0.03) & 0.94 (0.09) \\ 
				&   & $\beta_3$ & 0.97 (0.11) & 0.65 (0.25) & 0.78 (0.11) & 0.95 (0.09) & 0.86 (0.07) & 0.94 (0.13) \\ 
				&&&&&&&&\\[-2ex]
				&   & $\beta_1$ & 0.99 (0.01) & 0.73 (0.18) & 0.98 (0.02) & 0.96 (0.08) & 1.00 (0.01) & 0.87 (0.11) \\ 
				& 5000 & $\beta_2$ & 1.00 (0.01) & 0.55 (0.16) & 0.98 (0.02) & 0.92 (0.10) & 1.00 (0.01) & 0.89 (0.10) \\ 
				&   & $\beta_3$ & 0.96 (0.06) & 0.66 (0.30) & 0.89 (0.06) & 0.97 (0.08) & 0.90 (0.05) & 0.96 (0.11) \\ 
				\hline
			\end{tabular}
			\begin{tablenotes}
				\item STTV: the soft-thresholding time-varying effects  Cox model; RegTV: the regular time-varying effects  Cox model.
			\end{tablenotes}
		\end{threeparttable}
	\end{table}

	We set a total of 100 grid points on $[0,3]$, counting the number of $t_g$ in each set as its cardinality. Table \ref{ch3:tab:TPRTNR} shows that the soft-thresholding time-varying effects  Cox model has a higher inference-based true negative ratio than the regular time-varying effects  Cox model. Although the inference-based true positive and negative ratios are more reliable with controlled false discovery rates, their computational burden increases when the sample size increases. Therefore, the estimation-based true positive and negative ratios are favorable for large datasets as their calculation merely depends on the estimations. First, our method presents a better  estimation-based true  negative ratio ratio, indicating   our method can detect  zero-effect regions well. Second,  as documented in Table \ref{ch3:tab:TPRTNR}, our 
	method also presents a higher estimation-based true positive ratio than the inference-based true positive ratio, indicating a better performance of our method in inferring non-zero effects.

	\section{Analysis of the Boston Lung Cancer Survivor Cohort}\label{ch3:sec:real}
	We apply our method to study a subset of the Boston Lung Cancer Survivor Cohort (BLCSC) \cite{Christiani2017}. 
	\yl{The data consist of $n=599$ individuals, among whom $148$ (24.7\%) were alive and $451$ (75.3\%) were dead by the end of the follow up. The primary endpoint was overall survival measuring the time lag from the diagnosis of lung cancer to death or the end of the study, which ever came first. 
	 The range of the observed survival time was from 6 days to $8584$ days, and the restricted mean survival and censoring times at $\tau=8584$ days were 2124 (SE: 105) and 4397 (SE: 187) days, respectively. The observed survival time was skewed to the right.} 
	Patients who were alive were younger than those of those who died (average age in years: 55.4 vs. 61.2), and were slightly less likely to  be  Caucasian (89.9\% vs. 95.8\%). With early-stage lung cancer including stages lower than II, e.g., 1A, 1B, IIA, and IIB, 64.2\% of the alive patients  had early-stage lung cancer, slightly higher than those who died  (62.3\%). The percentage of the alive patients who had surgery was 83.8\%, higher than that of the dead patients (63.0\%). See Table \ref{ch3:tab:smry} for more details.
	
	\begin{table}[!htb]
		\centering
		\begin{threeparttable}
			\caption{Summary of the patient characteristics}\label{ch3:tab:smry}
			\begin{tabular}{lll}
				\hline
				   & Alive & Dead \\ 
				&$(n=148)$ &$(n=451)$\\
				\hline
				&&\\[-1ex]
				Age & 55.4 (10.1) & 61.2 (10.8) \\ 
				Pack years& 34.4 (29.7) & 51.6 (38.5) \\ 
				Race &&\\
				\quad White (ref) & 133 (89.9\%) & 432 (95.8\%) \\ 
				\quad Others &  15 (10.1\%) & 19 (4.2\%) \\ 
				Education &&\\
				\quad Under high school (ref) & 10 (6.8\%) & 72 (16\%) \\ 
				\quad High school graduate & 30 (20.3\%) & 113 (25.1\%) \\ 
				\quad Above high school & 108 (73.0\%) & 266 (59.0\%) \\
				Sex&&\\
				\quad Female (ref) & 113 (76.4\%) & 256 (56.8\%) \\ 
				\quad Male & 35 (23.6\%) & 195 (43.2\%) \\ 
				Smoking status&&\\
				\quad Ever or never (ref) & 96 (64.9\%)&281 (62.3\%)\\
				\quad Current & 52 (35.1\%) & 170 (37.7\%) \\ 
				Cancer stage&&\\
				\quad Early (ref) &95 (64.2\%)&190 (42.1\%)\\
				\quad Late & 53 (35.8\%) & 261 (57.9\%) \\  
				Surgery & 124 (83.8\%) & 284 (63.0\%) \\ 
				Chemotherapy &48 (32.4\%) & 206 (45.7\%) \\ 
				Radiotherapy& 35 (23.6\%) & 184 (40.8\%) \\ [1ex]
				\hline
			\end{tabular}
			\begin{tablenotes}
				\small
				\item Continuous variables are presented in mean (standard deviation), and categorical variables are presented in count (percentage). Due to rounding, some summations of percentages for one variable are not one. Reference groups are marked.
			\end{tablenotes}
		\end{threeparttable}
	\end{table}
	
\begin{figure}[!htbp]
		\small
		\centering
		\captionsetup{width=0.9\textwidth}
		\includegraphics[width=.35\textwidth]{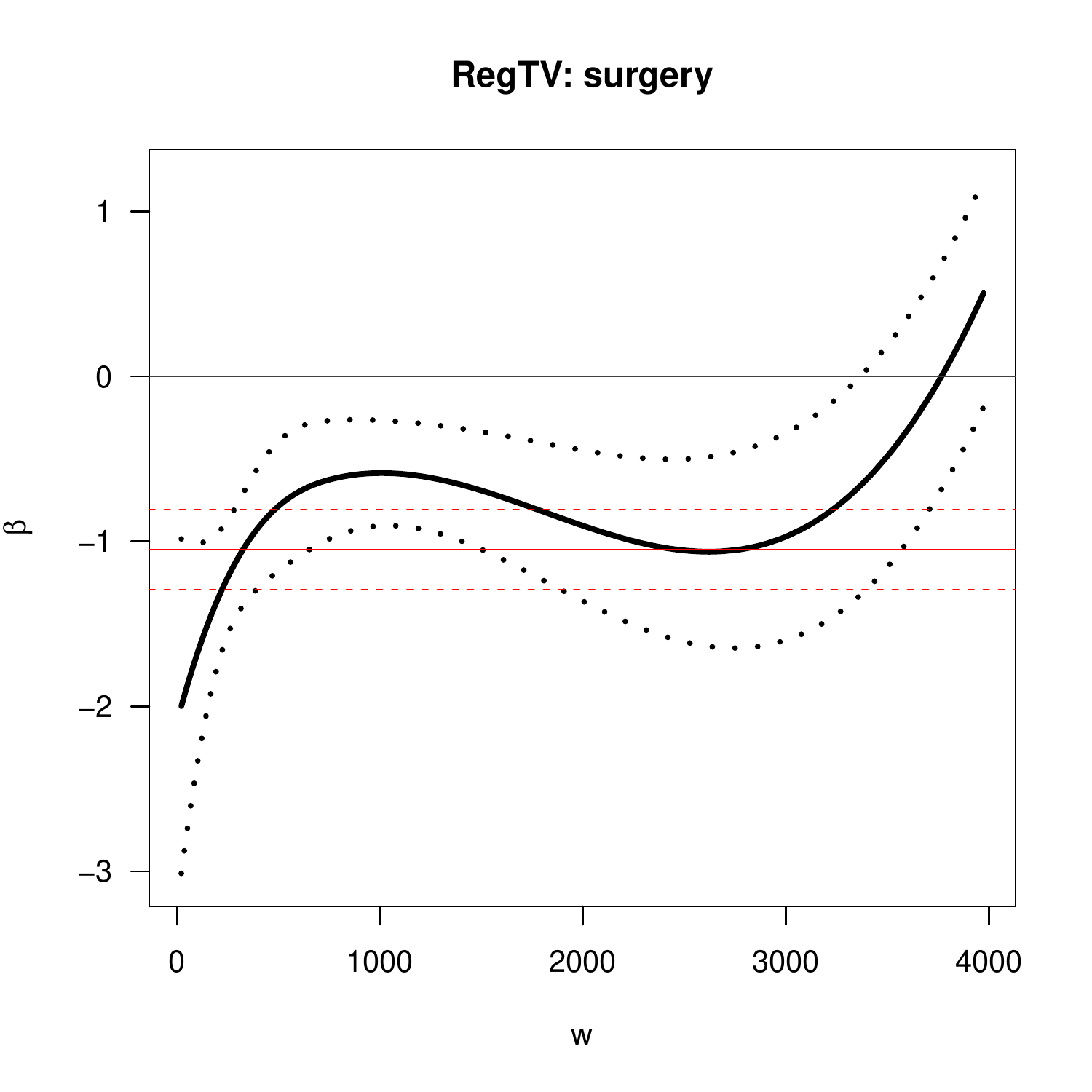}
		\includegraphics[width=.35\textwidth]{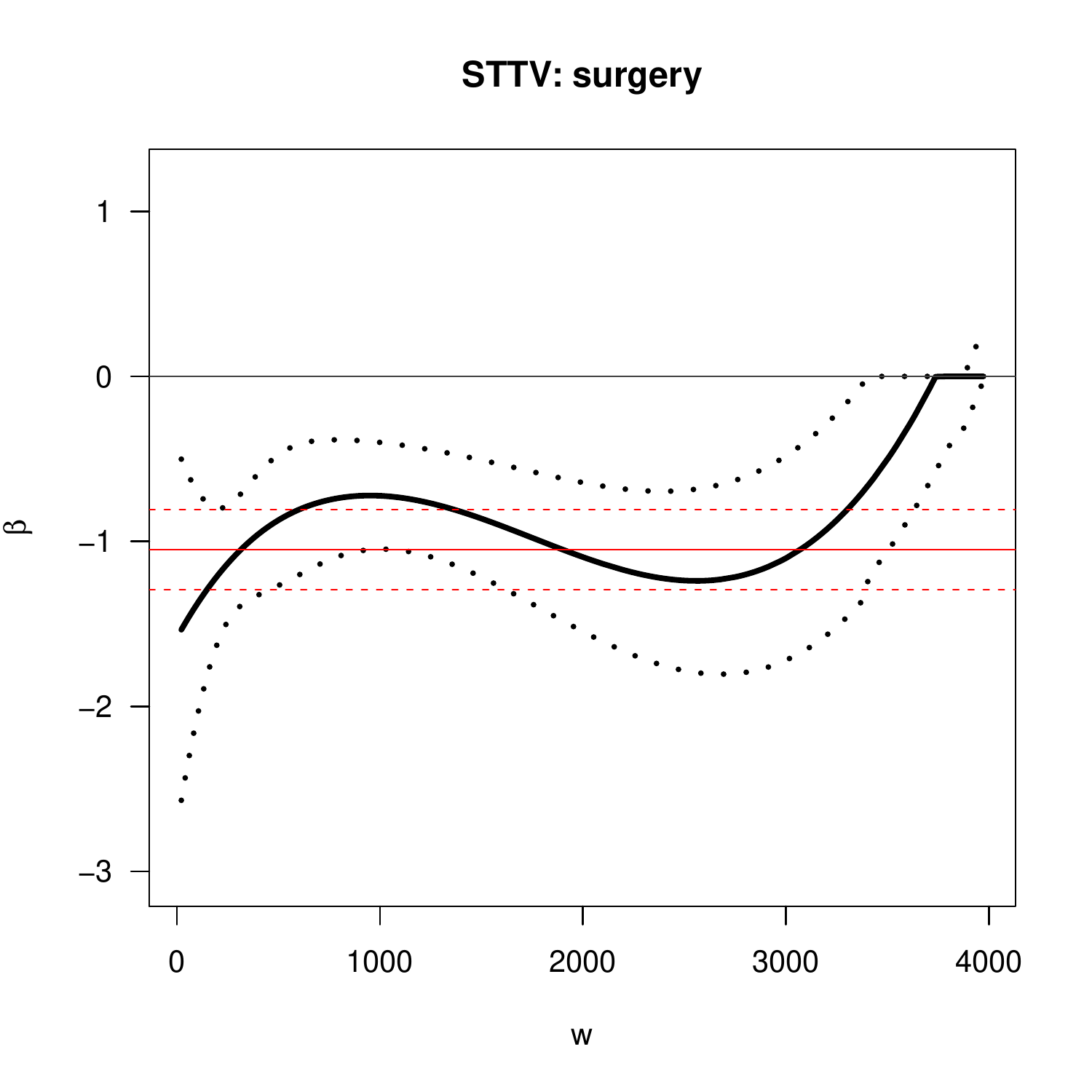}\quad
		\includegraphics[width=.35\textwidth]{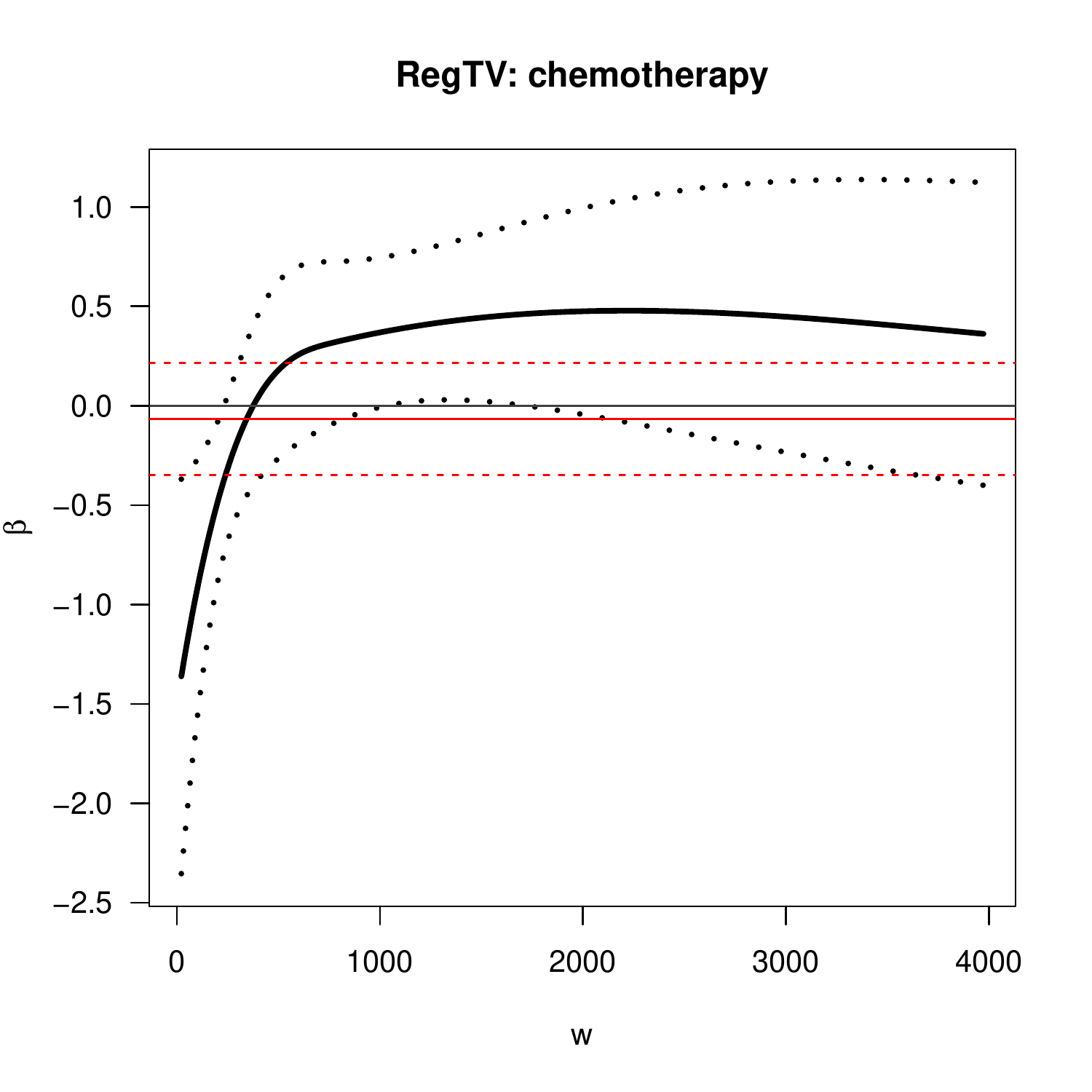}
		\includegraphics[width=.35\textwidth]{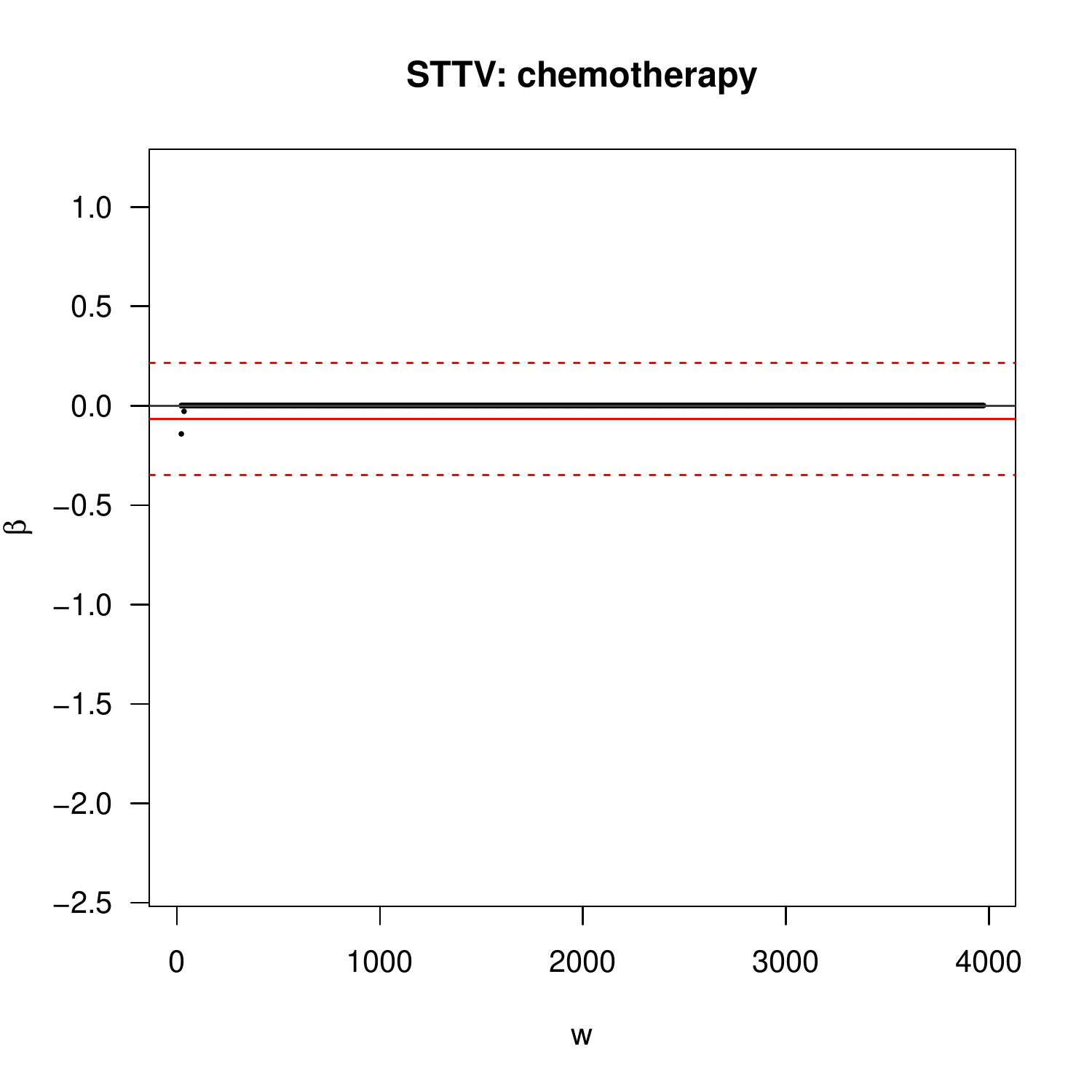}\quad
		\includegraphics[width=.35\textwidth]{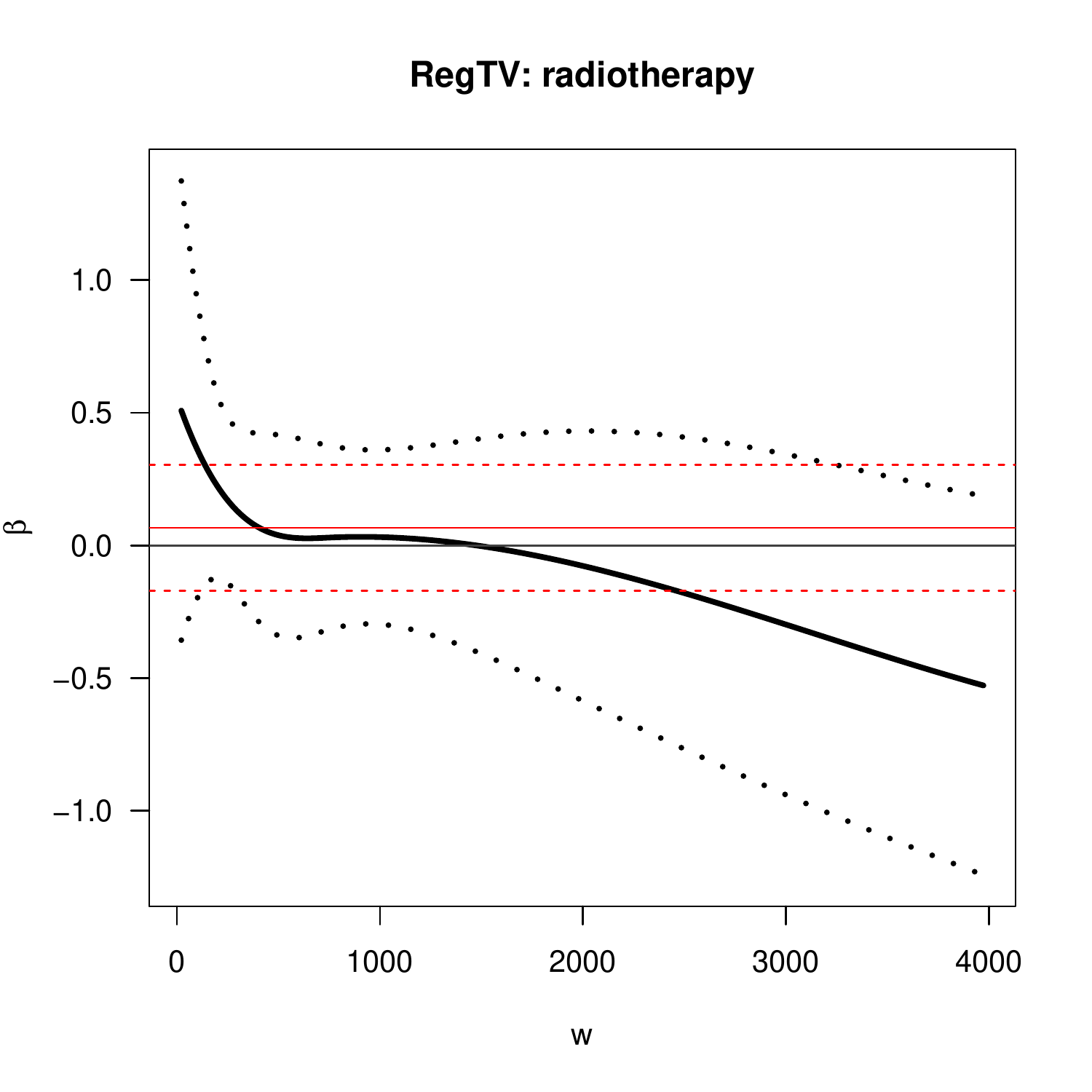}
		\includegraphics[width=.35\textwidth]{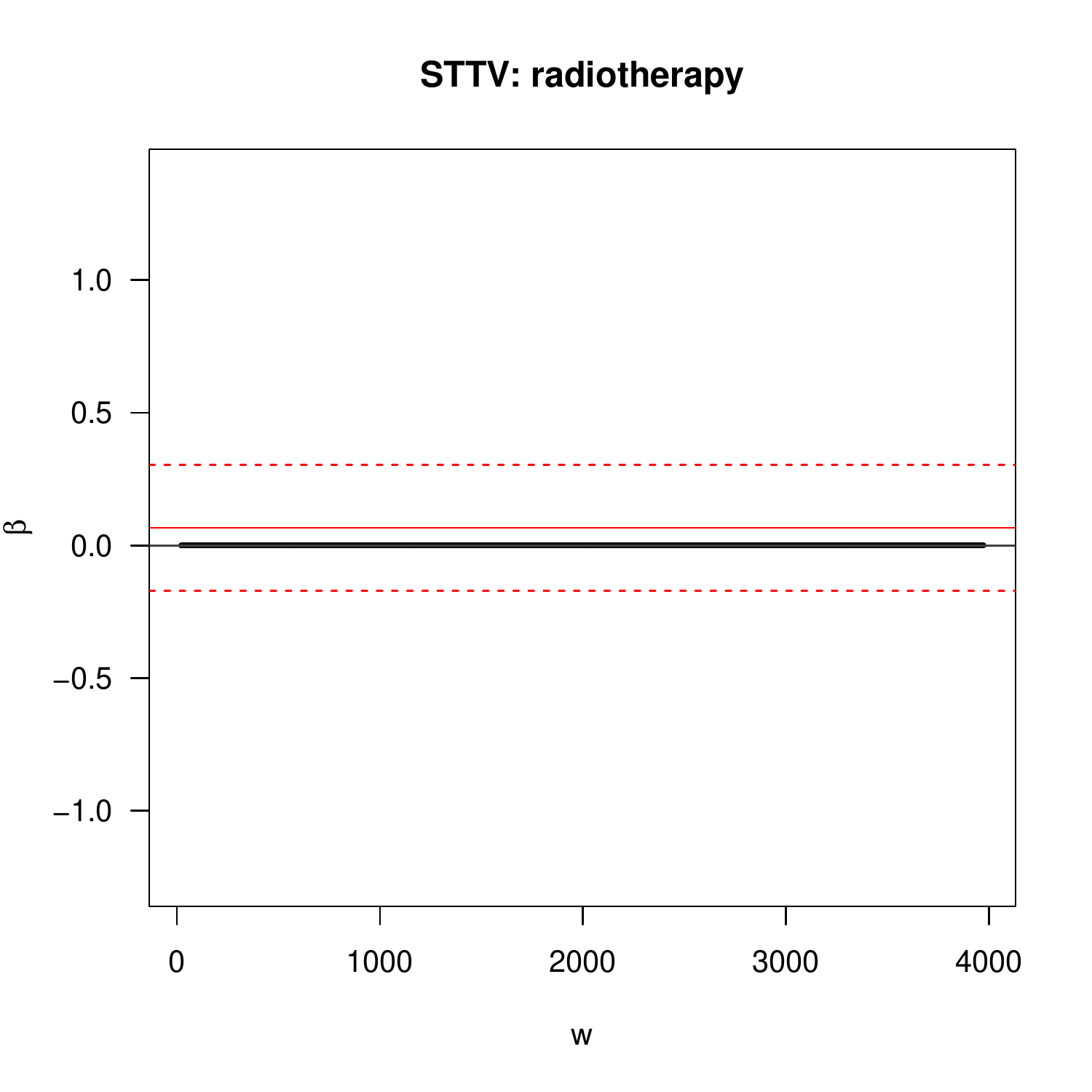}\quad
		\caption{Estimation results (part I) for the BLCSC data using the regular time-varying effects  Cox model (RegTV) and the soft-thresholding time-varying effects  Cox model (STTV): the solid lines are the estimated coefficient function curves; the dotted lines are the pointwise (sparse) confidence intervals; black lines are from varying coefficient models; red lines are from the constant effect Cox model.}
		\label{ch3:fig:real1}
	\end{figure}

\begin{figure}[!htbp]
		\small
		\centering
		\captionsetup{width=0.9\textwidth}
		\includegraphics[width=.35\textwidth]{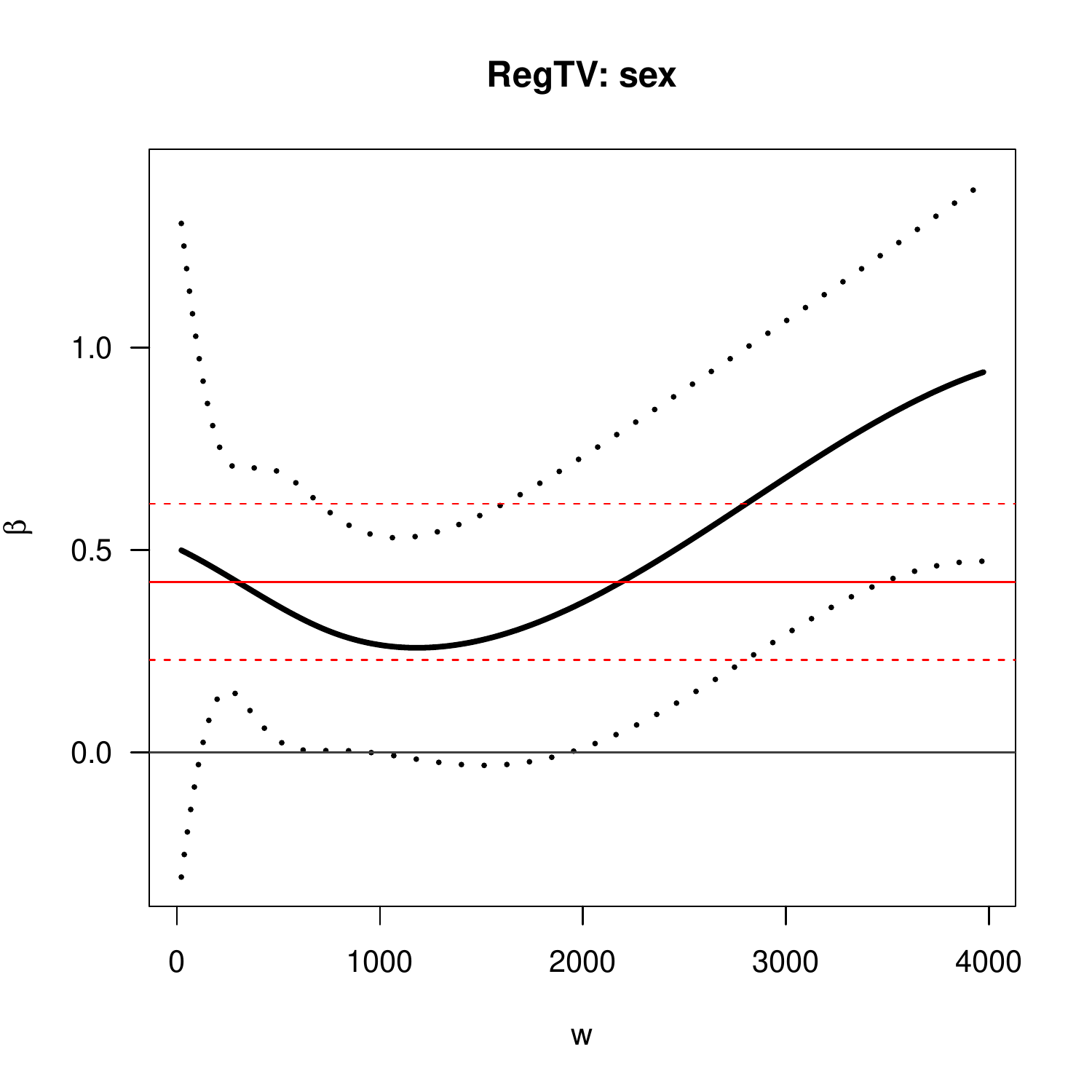}
		\includegraphics[width=.35\textwidth]{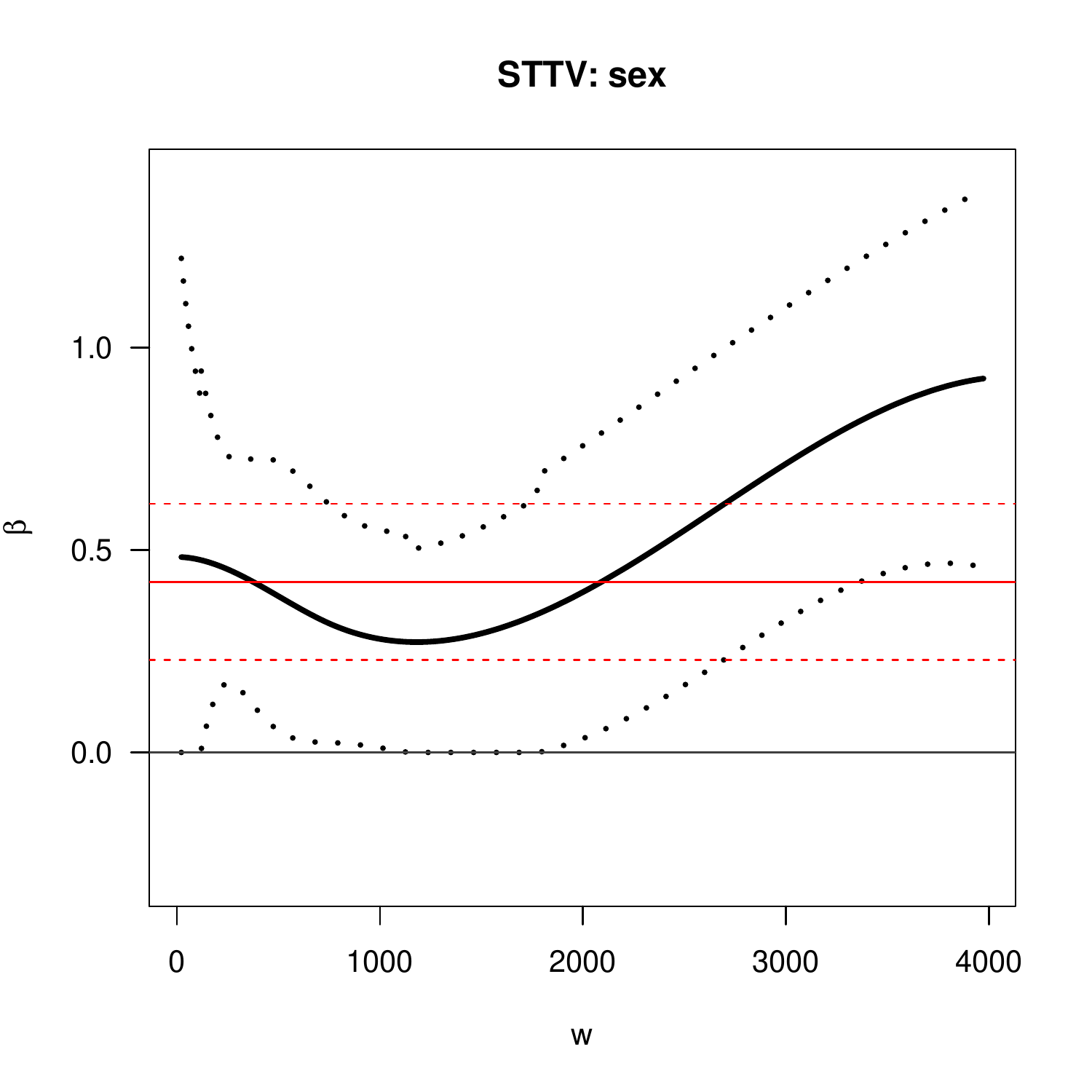}\quad
		\includegraphics[width=.35\textwidth]{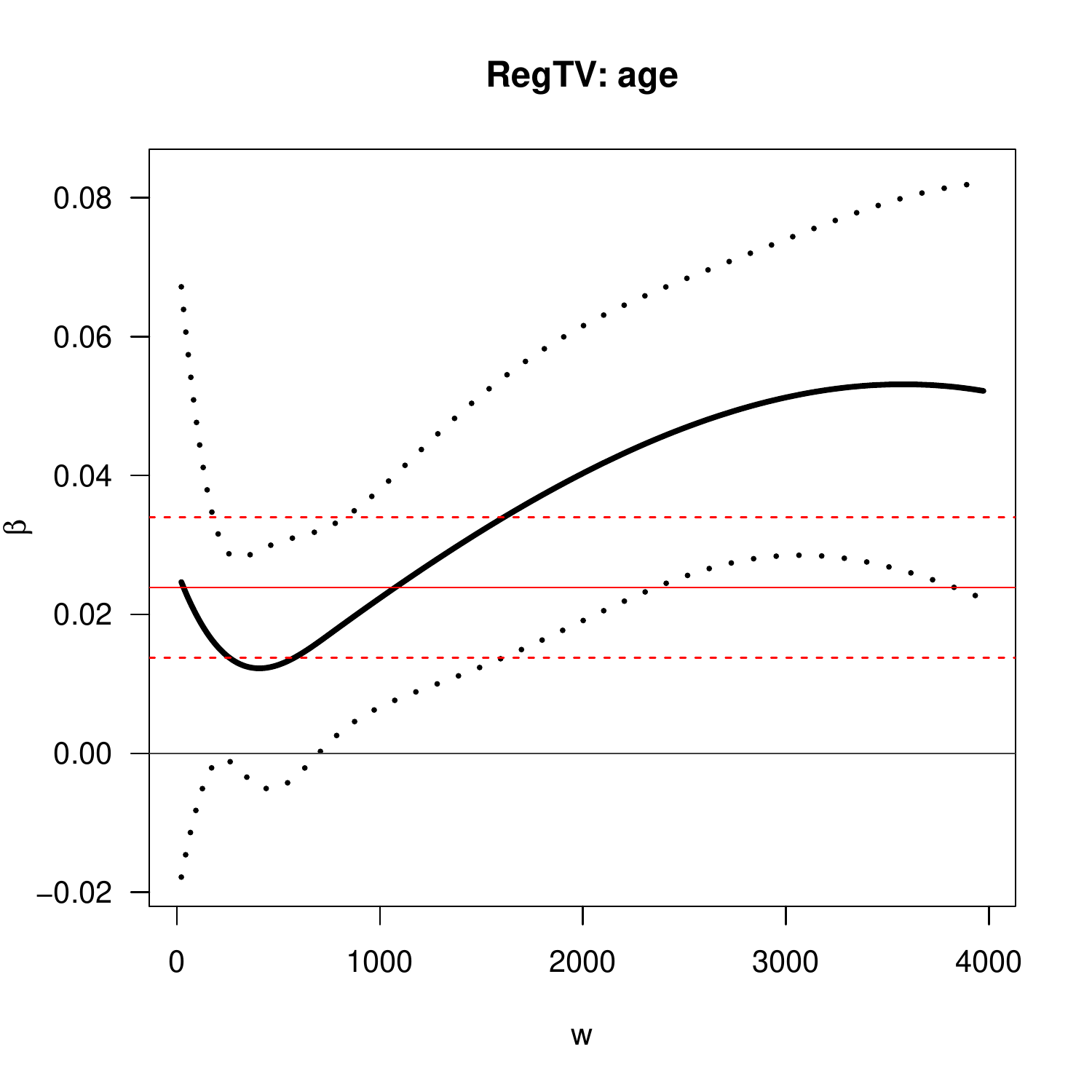}
		\includegraphics[width=.35\textwidth]{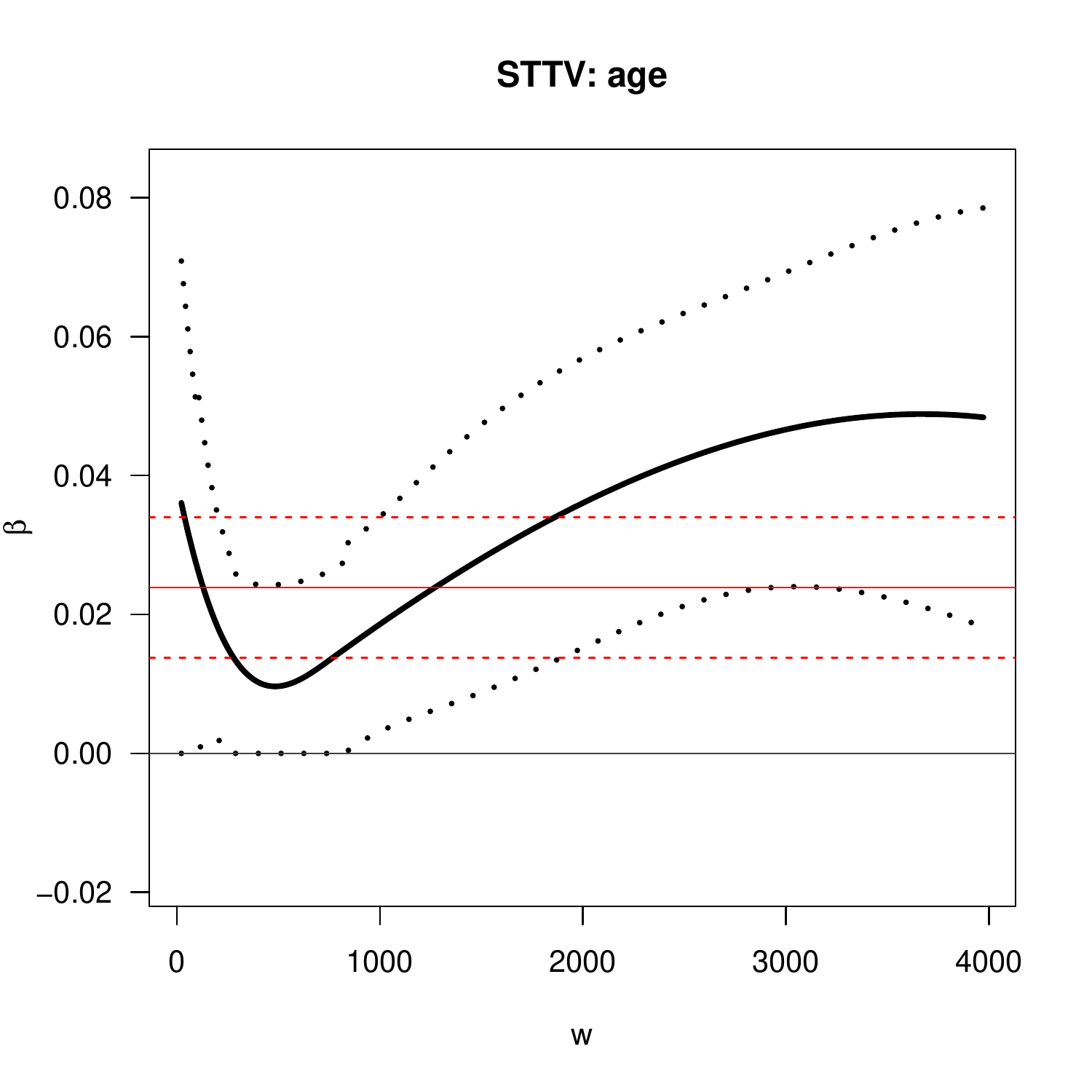}\quad
		\includegraphics[width=.35\textwidth]{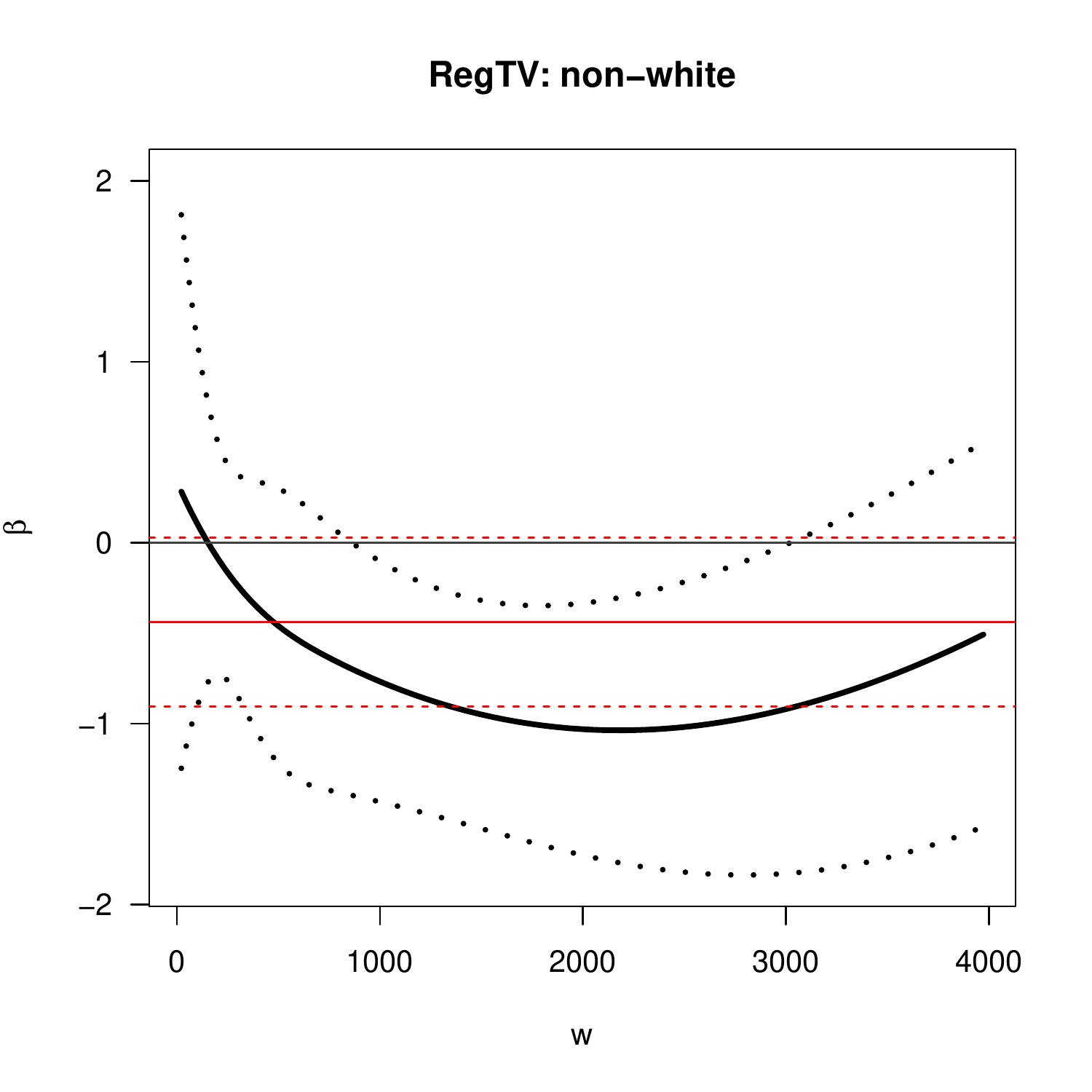}
		\includegraphics[width=.35\textwidth]{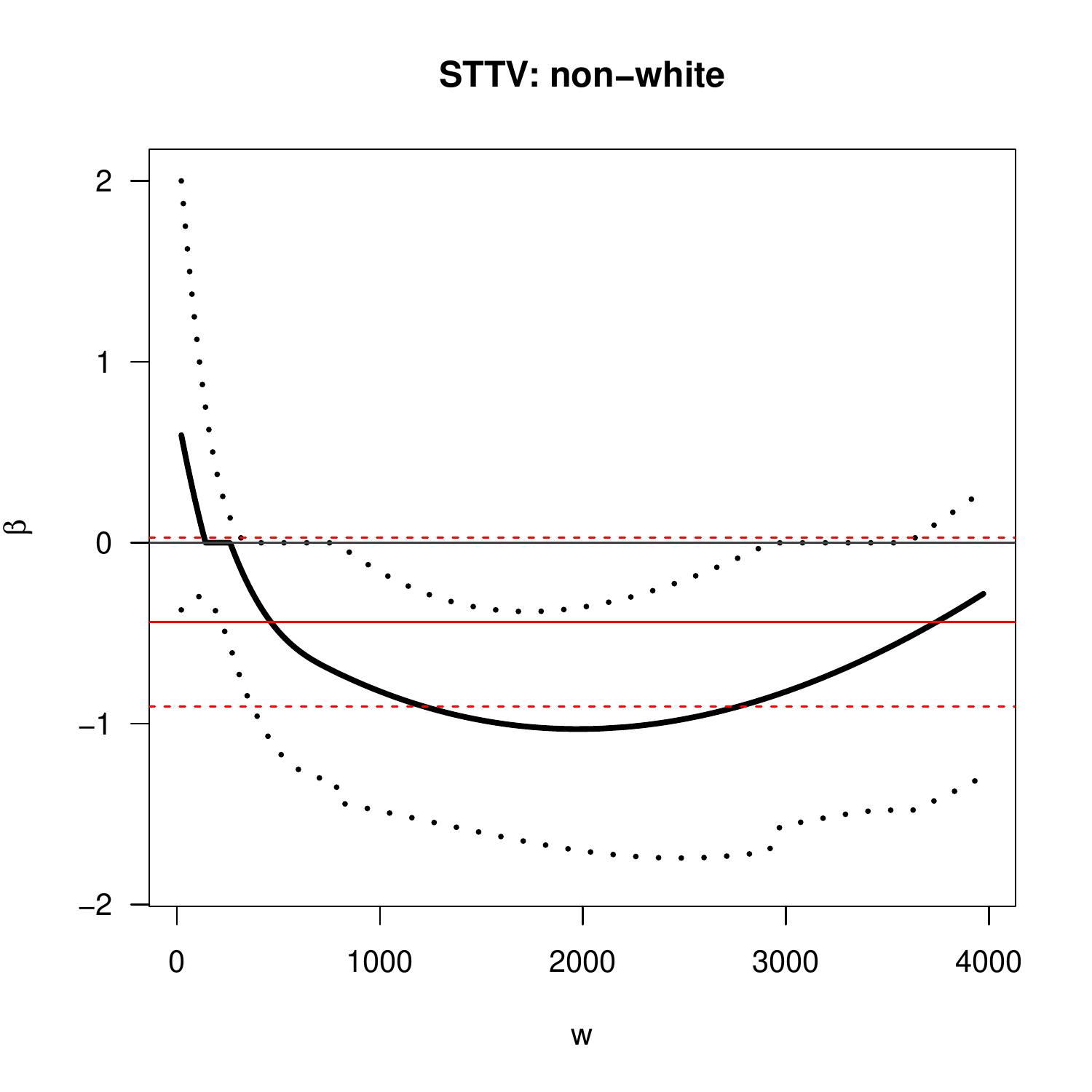}\quad
		\includegraphics[width=.35\textwidth]{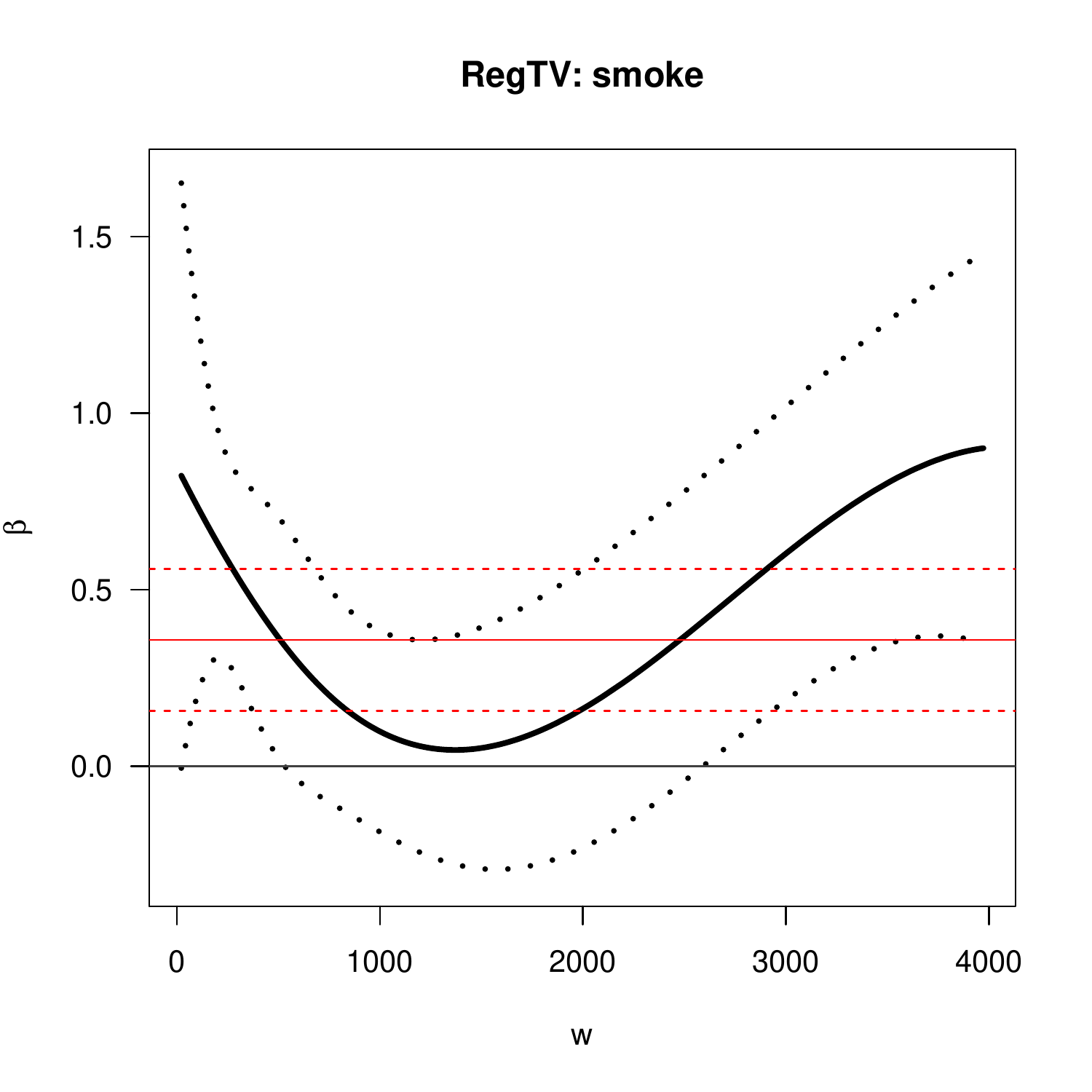}
		\includegraphics[width=.35\textwidth]{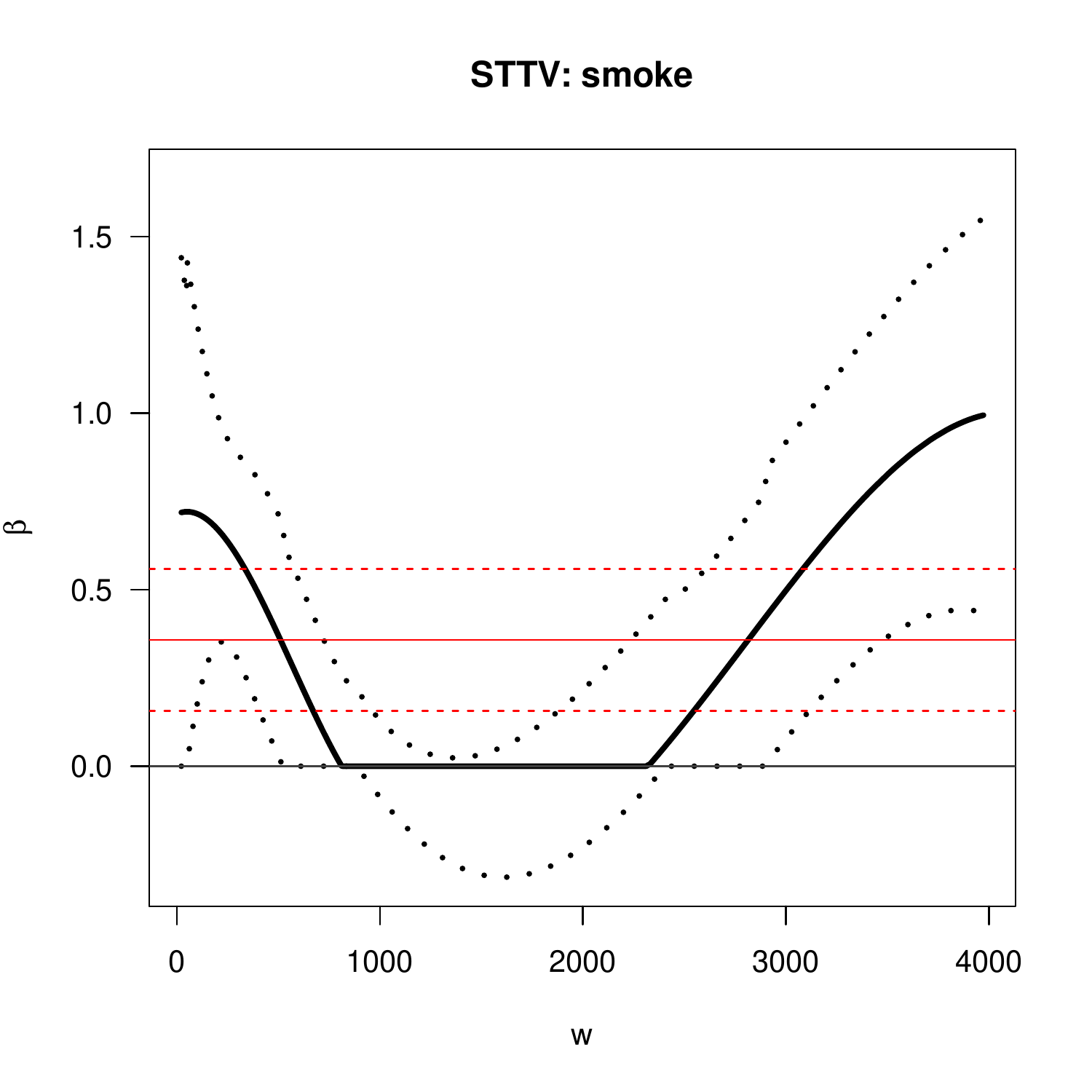}\quad
		\caption{Estimation results (part II) for the BLCSC data using the regular time-varying effects  Cox model (RegTV) and the soft-thresholding time-varying effects  Cox model (STTV): the solid lines are the estimated coefficient function curves; the dotted lines are the pointwise (sparse) confidence intervals; the black lines are from varying coefficient models and the red lines are from the Cox proportional hazards model.}
		\label{ch3:fig:real2}
	\end{figure}

\begin{figure}[!htbp]
		\small
		\centering
		\captionsetup{width=0.9\textwidth}
		\includegraphics[width=.35\textwidth]{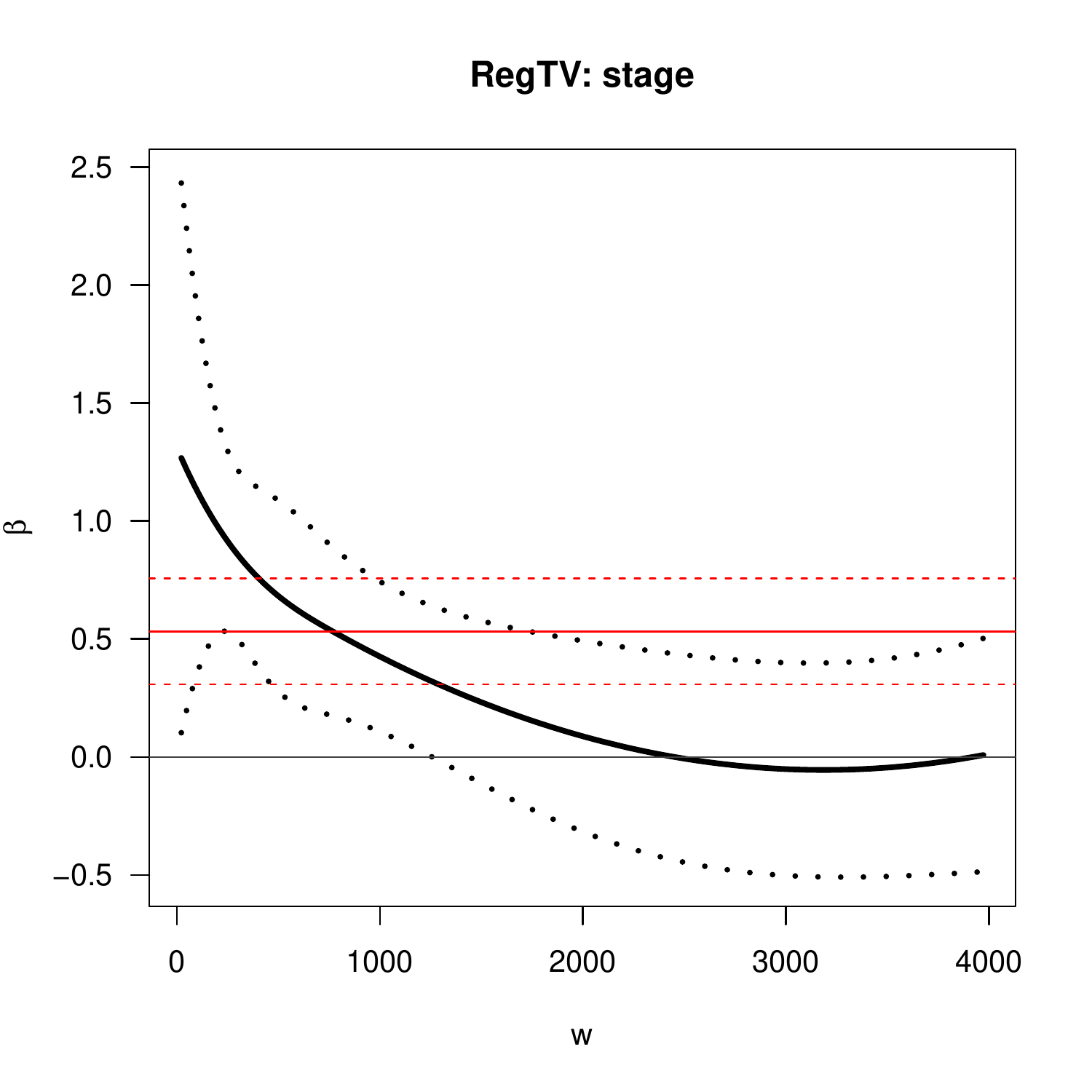}
		\includegraphics[width=.35\textwidth]{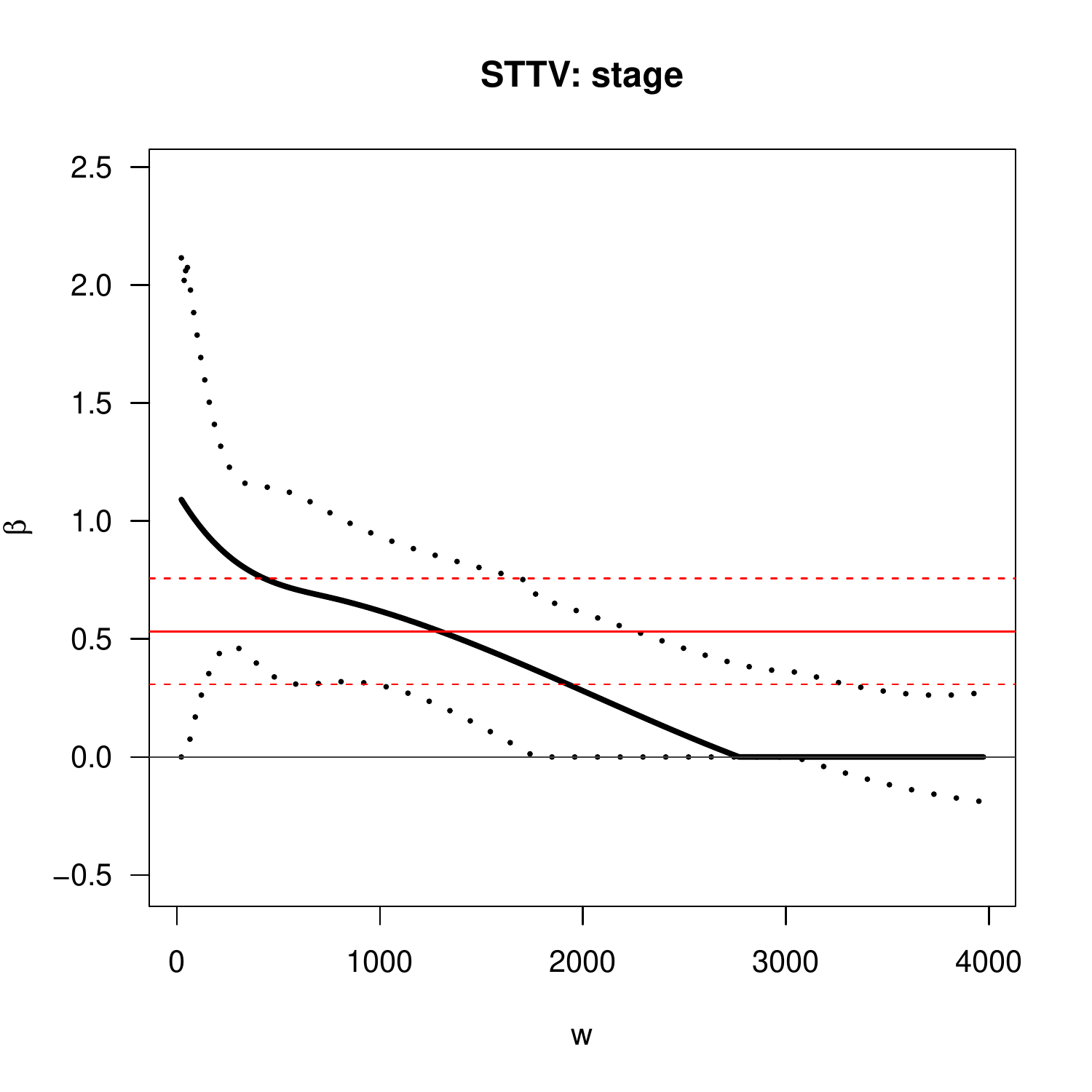}\quad
		\includegraphics[width=.35\textwidth]{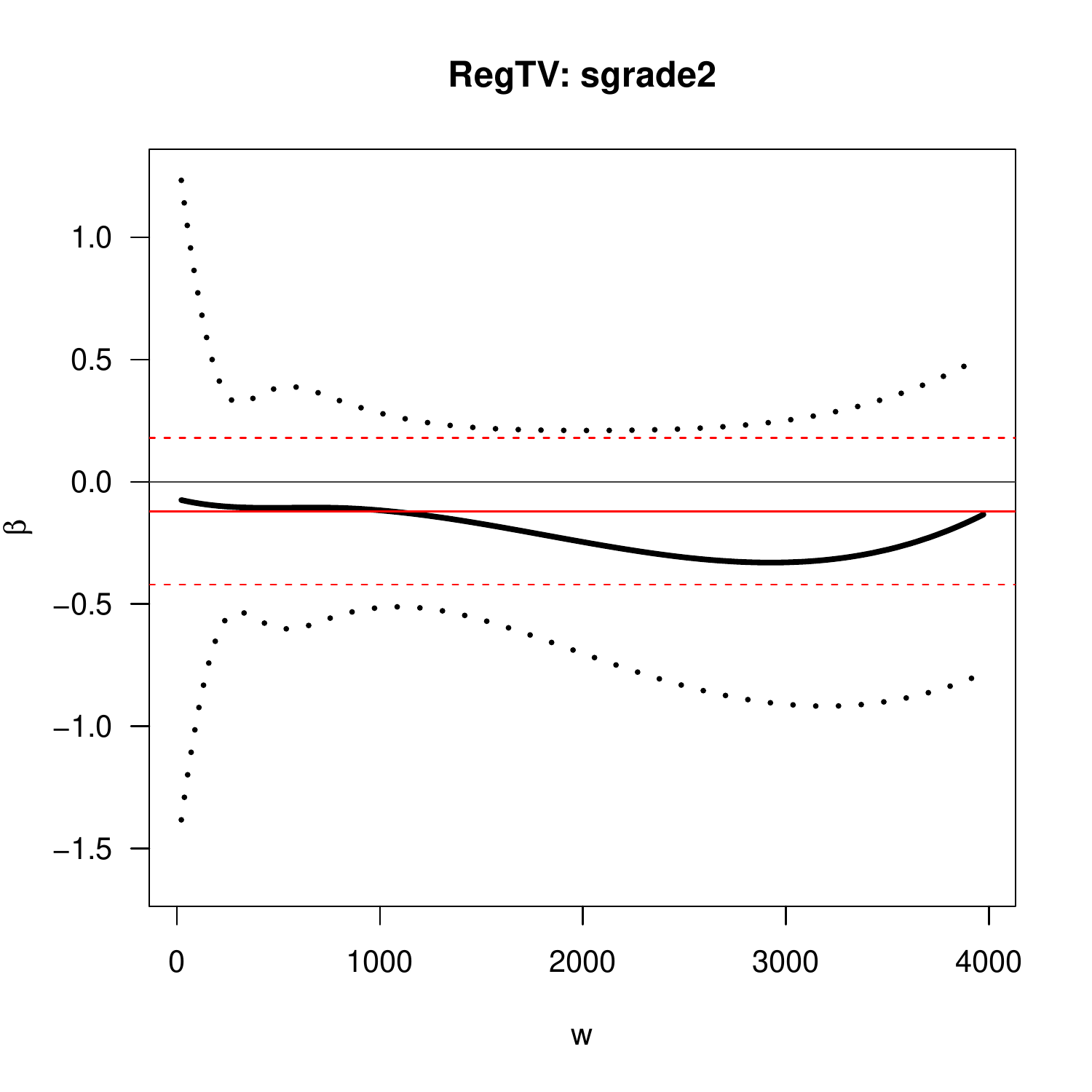}
		\includegraphics[width=.35\textwidth]{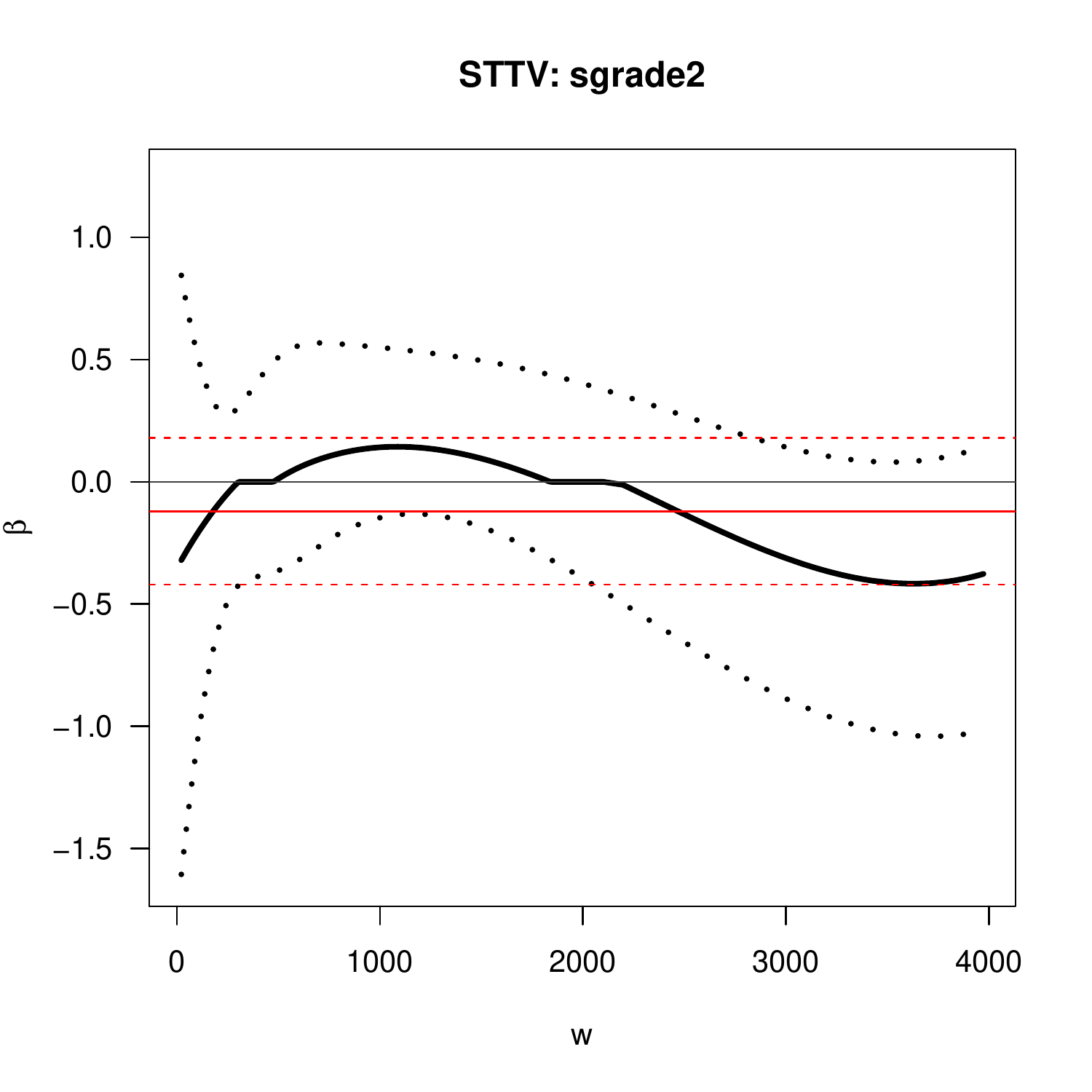}\quad
		\includegraphics[width=.35\textwidth]{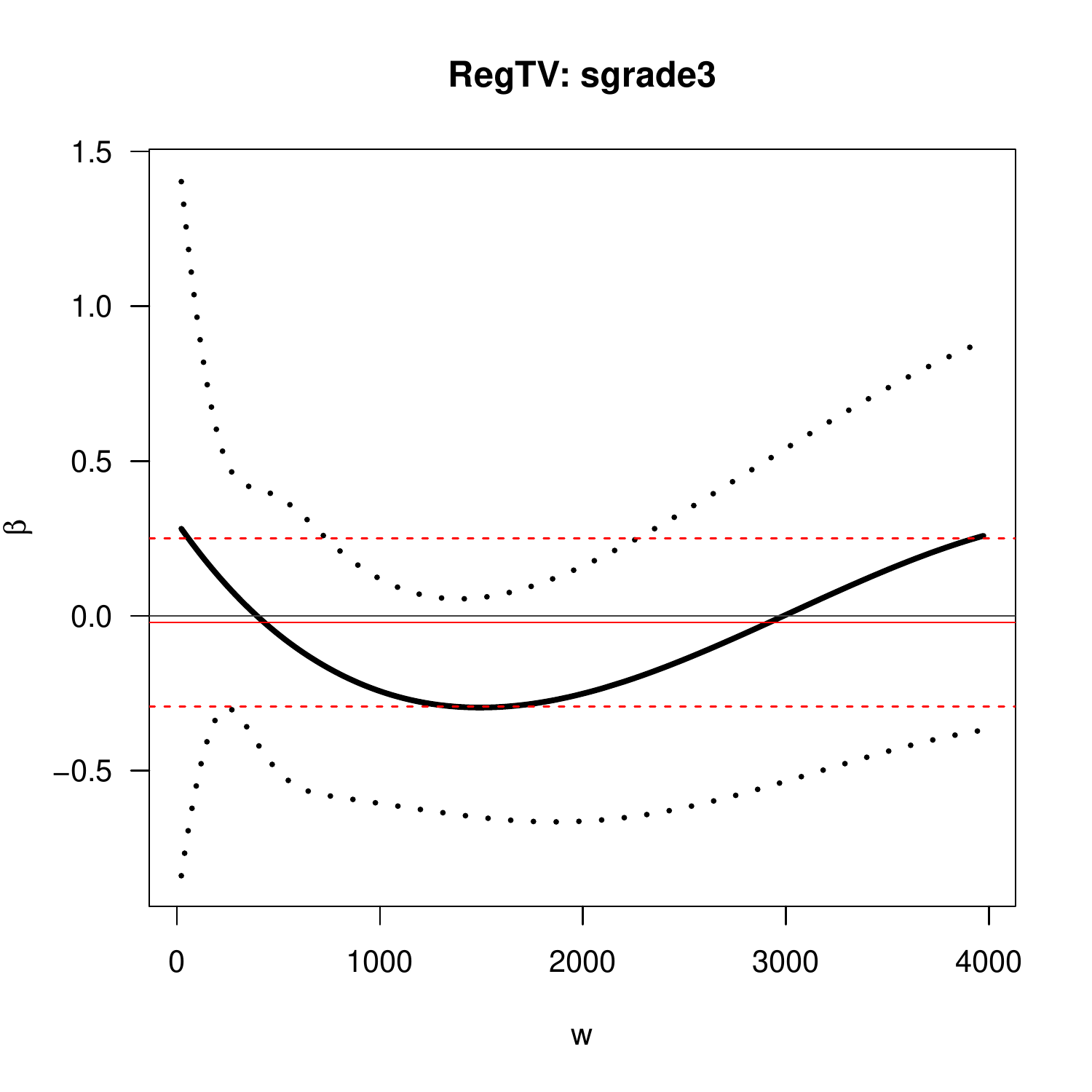}
		\includegraphics[width=.35\textwidth]{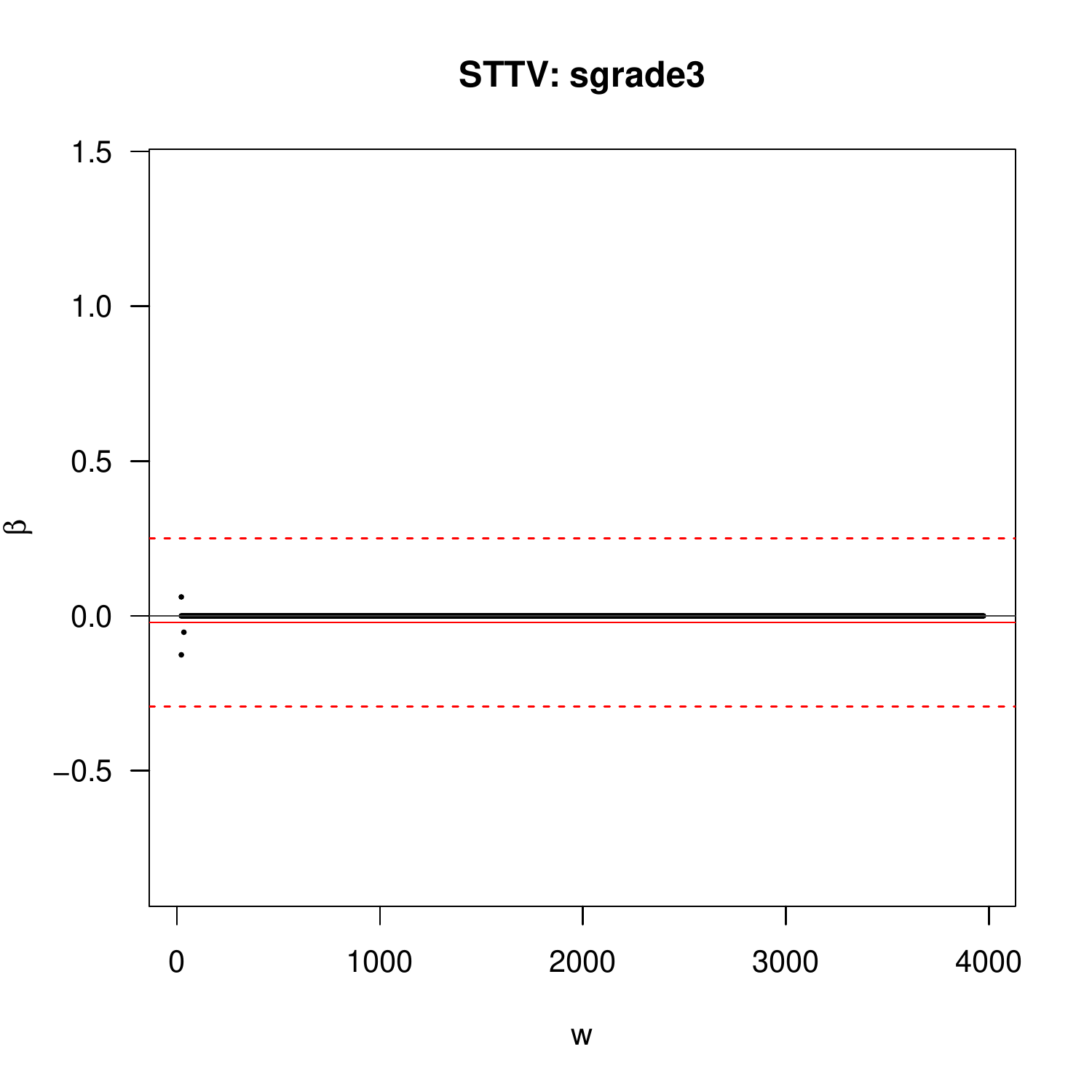}\quad
		\caption{Estimation results (part III) for the BLCSC data using the regular time-varying effects  Cox model (RegTV) and the soft-thresholding time-varying effects  Cox model (STTV): the solid lines are the estimated coefficient function curves; the dotted lines are the pointwise (sparse) confidence intervals; the black lines are from varying coefficient models and the red lines are from the  Cox proportional hazards model.}
		\label{ch3:fig:real3}
	\end{figure}

	Included in our analysis are  age, race, education, sex, smoking status, cancer stage, and treatments received (surgery, chemotherapy, and radiotherapy). For comparisons, we  fit the  
	data by using the Cox proportional hazards model, the regular time-varying effects Cox model (RegTV) and the soft-thresholding time-varying effects  Cox model (STTV).
	When implementing STTV, we set the needed parameters such as  $\rho, \alpha_j$
	and $K$ as in done in the simulation section. \yl{ In particular, with respect to the choice of $K$ and $\rho$, we have determined that $K=5$  by minimizing a 10-fold cross-validation error over a candidate set of  \{3, 5, 9, 13, 17, 21\}, while setting the penalty parameter $\rho$ to be $1/n^2$. }
 	We fit RegTV by using the penalized B-spline approach of  \cite{Zucker1990, Yan2012}. See the results as summarized in Figures \ref{ch3:fig:real1}, \ref{ch3:fig:real2} and \ref{ch3:fig:real3}.
	
	\yl{Compared with the regular time-varying effects  Cox model, the soft-thresholding time-varying effects  Cox model agrees more to  the  Cox proportional hazards model.} For some non-significant coefficients in the constant effect Cox model, STTV estimates those to be all zero over the time, such as for chemotherapy and radiotherapy. The results seem to be reasonable: insofar as  surgery had a strong protective effect for this group of lung cancer patients, adding chemotherapy or radiotherapy did not seem to be associated with additional protective or harmful impacts on  lung cancer patients' survival. This is consistent with the clinical practice that
	surgery is often the first line therapy for operable lung cancer patients \citep{zheng2018upfront}. Interestingly,  smoking at diagnosis was associated with higher short-term	(in the first 3 years post-diagnosis) and long-term (after 7 years post-diagnosis) mortality, but was not significantly associated with mortality between 3 and 7 years post-diagnosis, possibly a stabilization period for patients. The result highlights the importance of early cessation of smoking \citep{barbeau2006results}.  	
	
	The other results are equally interesting. Adjusting for all the other factors, the expected hazard was significantly higher among male patients than female patients; older patients had a significantly higher hazard than younger patients; non-white patients had a lower hazard than white patients;  a later cancer stage was strongly associated with worse lung cancer mortality. However, there were no significant associations between education levels and  lung cancer mortality. \yl{In conclusion, the results of STTV are consistent with those obtained by using the Cox model, but STTV can more accurately capture the time-varying effect of each factor. }
	

	\section{Discussion}\label{ch3:sec:discussion}
	
	To address the challenge of modeling time-varying coefficients with zero-effect regions in survival analysis,  we proposed a new soft-thresholding time-varying coefficient model, where the varying coefficients are piecewise smooth with zero-effect regions. To quantify  uncertainty of the estimates, we have designed a new type of sparse confidence intervals, which extend classical confidence intervals by accommodating exact zero estimates. Our  framework enables us to estimate non-zero time-varying effects and detect zero-effect regions simultaneously,  extending the already-widely-used  Cox models to a new 
	territory. The work pays tribute to Sir D.R. Cox, whose work has fundamentally influenced modern biomedical research.

\section*{Acknowledgements}
We thank Dr. David Christiani for providing the BLCSC data. \yl{We thank two referees for their insightful comments that have improved the presentation and the quality of the submission.} The work is partially supported by  grants from NIH (R01CA249096 and U01CA209414). 

\section*{Appendix}
\addcontentsline{toc}{section}{Appendix}
\noindent\textbf{Proof of Theorem \ref{ch3:thm:converg}}:

For every $f \in \mF_0$, by Corollary 6.21 of Schumaker (2007) \cite{Schumaker2007} , there exists an $f_n\in \mF$, $||f_n-f||_{\infty} = O(q^{-m})$. For any $\delta_1 >0$ and $ \delta_2>0$, there exists an $\eta$ (when constructing $h$) such that $| |h(f_n)-\zeta(f_n)||<\delta_1$ and $||h(f)-\zeta(f)||<\delta_2$. Then we have,
\begin{equation*}
\begin{split}
||h(f_n)-h(f)|| & < ||h(f_n)-\zeta(f_n)|| + ||\zeta(f_n)-\zeta(f)||+||\zeta(f)-h(f)||\\
& = A_1 + A_2 +A_3.
\end{split}
\end{equation*}

Let $\delta_1=O(q^{-m})$ and $\delta_2=O(q^{-m})$, then $A_1 < \delta_1 = O(q^{-m})$ and $A_2 < \delta_2 = O(q^{-m})$. We have $A_3 \leq ||f_n-f|| = O(q^{-m})$ because the Lipschitz continuous property in Lemma 1 of Kang et al. (2018) \cite{Kang2018}.  Therefore, $||h(f_n)-h(f)||_{\infty} = O(q^{-m})$. For simplicity of notation, let $h_{nj}$ denote $h(f_{nj})$ and $h_{0j}$ denote $h(f_{0j})$.  

Let $g_n = \sum_{j=1}^{p} Z_jh_{nj}$.  Then, given $\bZ$, $|g_n-g_0| = |\sum_{j=1}^{p} Z_j(h_{nj}-h_{0j})| \leq  \sum_{j=1}^{p} |Z_j| |(h_{nj}-h_{0j})| = O_p(q^{-m})$. Thus, we have $||g_n-g_0|| =   O_p(q^{-m})$. 

By  Lemma 5.1 of Huang (1999) \cite{Huang1999}, $||\hat{g}_n-g_n||_2^2 = o_p(1)$. We then only need to prove
\begin{equation} \label{ch3:eq_phi}
\rm E \sup_{\delta/2 < ||g-g_n|| \leq \delta} |M_n(g)-M_n(g_n)-(M_0(g)-M_0(g_n))| = O_p(n^{-\frac{1}{2}}\delta (q^{\frac{1}{2}} + log^{\frac{1}{2}}(1/\delta))).
\end{equation} 
It follows that 
\begin{equation*}
\begin{split}
&M_n(g)-M_n(g_n)-\{M_0(g)-M_0(g_n)\}\\
=&P_{\Delta_n} m_n(\cdot,g) -P_{\Delta_n}m_n(\cdot,g_n)  - P_{\Delta}m_0(\cdot,g)+P_{\Delta}m_0(\cdot, g_n)\\
=& P_{\Delta_n} m_n(\cdot,g) - P_{\Delta} m_0(\cdot,g) -P_{\Delta_n} m_n(\cdot,g_n) +P_{\Delta} m_0(\cdot,g_n)\\
=&P_{\Delta_n} \{\log S_{0n}(\cdot,g)-\log S_{0}(\cdot,g) \} - P_{\Delta_n}\{\log S_{0}(\cdot,g_n)-\log S_{0}(\cdot,g_n)\} \\
= & J_{1n} +J_{2n}.
\end{split}
\end{equation*}

For any $\beta \in \mH_n$ and any $\alpha>0$, we can find at least one $f \in \mF_n$ such that $\beta =  \zeta(f,\alpha)$, then $\log N_{[]}(\epsilon,\mH_n, \delta) \leq \log N_{[]}(\epsilon,\mF_n, \delta)  \lesssim c_1q \log(\delta/\epsilon)$ by calculation in \cite{Shen1994}. Therefore, we can also obtain $\log N_{[]}(\epsilon,\mH_n, \delta) \lesssim c_2 q \log(\delta/\epsilon)$ according to its construction. Because both $\exp$ and $\log$ are monotone functions, we have $\log N_{[]}(\epsilon,E_{n,\delta}, \delta) \lesssim c_2 q \log(\delta/\epsilon) +  c_3 q \log(\delta/\epsilon) \lesssim c_4 q \log(\delta/\epsilon) $, where $c_4 = \max(c_2, c_3)$. 

Therefore, $J_{[]}(\delta, \epsilon_{n,\delta}, \rho) = \int_{0}^{\delta} \sqrt{1+ \log N_{[]}(\epsilon,E_{n,\delta}, \rho)} {\rm d}\epsilon \lesssim \delta q^{\frac{1}{2}}$. By Lemma 3.4.2 of Van Der Vaart and Wellner (1996) \cite{van1996weak}, we have
\begin{equation}
\rm E ||J_{1n}|| \lesssim n^{-\frac{1}{2}} q^{\frac{1}{2}} \delta (1+ \frac{q^{\frac{1}{2}}\delta}{\delta^2\sqrt{n}}c_5) = O(n^{-\frac{1}{2}} q^{\frac{1}{2}}\delta).
\end{equation}

On the other hand, we have
\begin{equation*}
\begin{split}
\sup_{ ||g-g_n|| \leq \delta} |J_{2n}| & \leq 2 \sup_{0\leq t \leq\tau, ||g-g_n|| \leq \delta} \left| log \frac{S_{0n}(\cdot,g)}{S_{0n}(\cdot,g_n)}-log\frac{S_0(\cdot, g)}{S_0(\cdot,g_n)} \right|\\
& \lesssim \sup_{0\leq t \leq\tau, ||g-g_n|| \leq \delta}   \left| \frac{S_{0n}(\cdot,g)}{S_{0n}(\cdot,g_n)}-\frac{S_0(\cdot, g)}{S_0(\cdot,g_n)} \right| \\
& \lesssim \sup_{0\leq t \leq\tau, ||g-g_n|| \leq \delta} \left| \frac{S_{0n}(\cdot,g)S_0(\cdot,g_n) - S_{0n}(\cdot,g_n)S_0(\cdot, g) }{S_{0n}(\cdot,g_n)S_0(\cdot,g_n)} \right|.
\end{split}
\end{equation*}

Since the denominator is bounded away from 0 with probability approaching to 1, we only need to consider the numerator. It follows that
\begin{equation*}
\begin{split}
& S_{0n}(\cdot,g)S_0(\cdot,g_n) - S_{0n}(\cdot,g_n)S_0(\cdot, g) \\
= & S_0(t,g_n)\{S_{0n}(t,g)-S_{0n}(t,g_n)-S_0(t,g)+ S_0(t,g_n)\} - \\     &\{  S_{0n}(t,g_n)-S_{0}(t,g_n)\}\{S_{0}(t,g)-S_{0n}(t,g)\} \\
= & I_{1n} - I_{2n}.
\end{split}
\end{equation*}

Since $I_{1n} = S_0(t,g_n){Y(t) [\exp(g(z))-\exp(g_n(z))]}$, we consider the class of function $Y(t)\exp(g(z))$. Since $\exp$ is monotone and the entropy of the class of indicator function $Y(t) = I[0\leq t \leq \tau ] $ is $\delta \log^{\frac{1}{2}} (1/\delta)$, we have that the entropy of the class of function $Y(t)exp(g(z))$ is $\delta (q^{\frac{1}{2}} + log^{\frac{1}{2}} (1/\delta)) $. By Lemma 3.4.2 of Van Der Vaart and Wellner (1996) \cite{van1996weak}, $I_{1n} \lesssim n^{-\frac{1}{2}}\delta (q^{\frac{1}{2}} + \log^{\frac{1}{2}}(1/\delta))$.

By Taylor's expansion and Jensen's inequality, we have
\begin{equation*}
\begin{split}
|S_0(t,g)-S_0(t,g_n)| &\leq E(Y(t)[\exp(g)-\exp(g_n)]) \\
& \leq \rm E(\exp(g_n)|g-g_n|) \\
&\lesssim (\rm E(g-g_n)^2)^{\frac{1}{2}} = O_p(\delta).
\end{split}
\end{equation*}
Since $S_n(t,g_n)-S_0(t,g_n)=O_p(n^{-\frac{1}{2}}q^{\frac{1}{2}})$, we obtain $I_{2n} = O_p(n^{-\frac{1}{2}}q^{\frac{1}{2}}\delta)$.

Therefore, $\sup_{ ||g-g_n|| \leq \delta} |J_{2n}| \lesssim n^{-\frac{1}{2}}\delta (q^{\frac{1}{2}} + \log^{\frac{1}{2}}(1/\delta))$. Thus, we have $M_n(g)-M_n(g_n)-\{M_0(g)-M_0(g_n)\} = O_p(n^{-\frac{1}{2}}\delta (q^{\frac{1}{2}} + \log^{\frac{1}{2}}(1/\delta)))$.

By Theorem 3.4.1 of Van Der Vaart and Wellner (1996) \cite{van1996weak}, the key function $\phi(\delta)$ takes the form of $\phi_n(\delta) = \delta(q^{\frac{1}{2}}+log^{\frac{1}{2}}(1/\delta))$. Therefore, $||(\hat{g}_n- g_n)||_2 = O_p((q/n)^{\frac{1}{2}})$. Therefore,  we have 
\begin{equation}
\begin{split}
||\hat{g}_n-g_0||_2^2  & \leq ||\hat{g}_n-g_n||_2^2+||g_n-g_0||_2^2 \\
& \leq O_p(q/n)+O_P(q^{-2m})  \\
& \leq O_p(r_n),
\end{split}
\end{equation}
where $r_n = q/n+q^{-2m}$.

Then by Lemma 1 of Stone (1985) \cite{Stone1985}, we have
\begin{equation}
E(Z_{j}\hat{h}_j(t)-Z_{j}h_j(t))^2 =  O_p(r_n), \quad 1\leq j \leq p.
\end{equation}

By Condition \ref{ch3:con:z}, there exists $ \delta, \epsilon>0 $, $\Pr(|Z_{j}|>\delta) > \epsilon$. Then
\begin{equation}
\begin{split}
E(Z_{j}\hat{h}_j(t)-Z_{j}h_j(t))^2 & >\Pr(|Z_{j}|>\delta)\delta^2 (\hat{h}_j(t) - h_j(t))^2 \\
& > \epsilon \delta^2 (\hat{h}_j(t) - h_j(t))^2.
\end{split}
\end{equation}

Therefore for any $t$, we have $(\hbeta_j(t) - \beta_j(t))^2 = O_p(r_n)$, i.e. $|\hbeta_j(t)-\beta_j(t)|= O_p(r_n^{1/2})$. Then we have $||\hbeta_j-\beta_j||_{\infty} = O_p(r_n^{1/2})$ for $j=1,\ldots,p$. $\blacksquare$

\hspace{2cm}

\noindent\textbf{Proof of Theorem \ref{ch3:thm_asymp}}:

We show Theorem  \ref{ch3:thm_asymp} is true when $\tau=1$. The extension to any $\tau < \infty$ satisfying condition C2 is straightforward and is omitted.

Following the counting process notation in Anderson and Gill (1982) \cite{Anderson1982}, we let 
\begin{equation*}
C(\bgam,t) = \sum_{i=1}^n \int_{0}^\top \sum_{j=1}^p Z_{ij}h_j(\bgam_j,s) {\rm d} N_i(s) -\int_0^\top \log \left\{ \sum_{i=1}^n Y_i(s) \exp\{\sum_{j=1}^p Z_{ij}(s)h_j(\bgam_j,s) \} \right\} {\rm d}\bar{N}(s),
\end{equation*}
then we have,
\begin{equation*}
{\rm PL}(\bgam) = C(\bgam,1) - \rho ||\btheta||_2^2.
\end{equation*}
Then for any $\bgam$,
\begin{equation*}
{\rm PL}^{'}(\bgam)  = C^{'}(\bgam, 1) - \rho \sum_{i=1}^{n} \btheta \otimes \bB(T_i).
\end{equation*}

By Taylor's expansion, we have that
\begin{equation*}
\{\rm PL\}^{'} (\hbgam) -  {\rm PL}^{'} (\tbgam)= \{\rm PL\}^{''} (\bgam^*) (\bgam-\tbgam),
\end{equation*}
where $\bgam^*$ is on the line segment between $\hbgam$ and $\tbgam$. Since $\{\rm PL\}^{'} (\hbgam)=0$, we have
\begin{equation*}
\begin{split}
\bgam-\tbgam &= -\left[ \{\rm PL\}^{''} (\bgam^*) \right]^{-1}   {\rm PL}^{'} (\tbgam)\\
& = -\left[ \{\rm PL\}^{''} (\bgam^*) \right]^{-1}  \left\{C^{'}(\tbgam,1) - \rho \sum_{i=1}^{n} \btheta_0 \otimes \bB(T_i) \right\} \\
& = -\left[ \{\rm PL\}^{''} (\bgam^*) \right]^{-1} C^{'}(\tbgam,1)  +  \rho  \left[ \{\rm PL\}^{''} (\bgam^*) \right]^{-1} \sum_{i=1}^{n} \btheta \otimes \bB(T_i).
\end{split}
\end{equation*}

The goal is to prove that for any non-zero $\ba$, 
\begin{equation*}
\frac{\ba^\top(\hbgam-\tbgam)}{\hsig(\ba)} \rightarrow_d N(0,1),
\end{equation*}
where $\hsig(\ba)= n\ba^\top \left[\{\rm PL\}^{''} (\tbgam) \right]^{-1}\Sigma(\tbgam,1)\left[ \{\rm PL\}^{''} (\tbgam) \right]^{-1}\ba$.

We claim that 
\begin{equation}\label{ch3:eq:part1}
\frac{\ba^\top \left[ -\{\rm PL\}^{''} (\bgam^*) \right]^{-1} C^{'}(\tbgam,1)}{\hsig(\ba)} \rightarrow_d N(0,1)
\end{equation}
and 
\begin{equation}\label{ch3:eq:part2}
\rho \ba^\top  \left[ \{\rm PL\}^{''} (\bgam^*) \right]^{-1} \sum_{i=1}^{n} \btheta \otimes \bB(T_i) /{\hsig(\ba)}  \rightarrow_p 0.
\end{equation}

To show \eqref{ch3:eq:part1}, we will utilize the martingale theories in Anderson and Gill (1982) \cite{Anderson1982} to prove that ${\ba^\top \left[ -\{\rm PL\}^{''} (\bgam^*) \right]^{-1} C^{'}(\tbgam,t)}/{\hsig(\ba)}$ is converging to a Gaussian process. 
Indeed, 
\begin{equation*}
C^{'}(\tbgam,t) = \sum_{i=1}^n \int_{0}^\top \left\{A_i(\tbgam,s)-E(\tbgam,s)\right\}  d M_i(s),
\end{equation*}
where $A_i(\tbgam,s)= \bU_i \otimes \bB_i$ and $E(\tbgam,s) = S_1(\tbgam,s)/S_0(\tbgam,s)$.
Then we have
\begin{equation*}
\frac{\ba^\top \left[ -\{\rm PL\}^{''} (\bgam^*) \right]^{-1} }{\hsig(\ba)} C^{'}(\tbgam,t) = \sum_{i=1}^n \int_{0}^\top \frac{\ba^\top \left[ -\{\rm PL\}^{''} (\bgam^*) \right]^{-1} }{\hsig(\ba)}\left\{A_i(\tbgam,s)-E(\tbgam,s)\right\}  d M_i(s).
\end{equation*}

Let $$H_i(s) = \frac{\ba^\top \left[ -\{\rm PL\}^{''} (\bgam^*) \right]^{-1} }{\hsig(\ba)} \left\{A_i(\tbgam,s)-E(\tbgam,s)\right\},$$ we then can show claim \ref{ch3:eq:part1} is true by applying Theorem I.2 in Anderson and Gill (1982) \cite{Anderson1982}. Condition (I.3) of Theorem I.2 is valid,  because by Conditions \ref{ch3:con:tau}, \ref{ch3:con:s012} and \ref{ch3:con:Sigma}, we have
\begin{equation*}
\begin{split}
\int_0^\top \sum_{i=1}^n H_i^2(s)\lambda_i(s) {\rm d}s =&  \ba^\top \left[ -\{\rm PL\}^{''} (\bgam^*) \right]^{-1} \cdot \\
& \int_0^\top  \sum_{i=1}^n \left\{A_i(\tbgam,s)-E(\tbgam,s)\right\}\left\{A_i(\tbgam,s)-E(\tbgam,s)\right\}^\top \lambda_i(s) {\rm d}s \cdot \\
& \left[ -\{\rm PL\}^{''}(\bgam^*) \right]^{-1} \ba/\hsig^2(\ba) \\
\rightarrow_p r(t),
\end{split}
\end{equation*}
where $r(t)$ is some positive function of $t$ and $r(1)=1$.

By similar arguments in Anderson and Gill (1982) \cite{Anderson1982}, condition (I.4) of Theorem I.2 is true by Conditions \ref{ch3:con:tau}, \ref{ch3:con:s012}, and \ref{ch3:con:Linde}. Then claim \eqref{ch3:eq:part1} is valid.

Claim \eqref{ch3:eq:part2} is valid because 
\begin{equation}
\rho \left|\ba^\top  \left[ \{\rm PL\}^{''} (\bgam^*) \right]^{-1} \sum_{i=1}^{n} \btheta \otimes \bB(T_i) /{\hsig(\ba)} \right| \leq O_p(n\rho )   \rightarrow_p 0
\end{equation}
by Condition \ref{ch3:con:rho}. Therefore, for any non-zero $\ba$, 
\begin{equation*}
\frac{\ba^\top(\hbgam-\tbgam)}{\hsig(\ba)} \rightarrow_d N(0,1),
\end{equation*}
where $\hsig(\ba)= n\ba^\top \left[\{\rm PL\}^{''} (\tbgam) \right]^{-1}\Sigma(\tbgam,1)\left[ \{\rm PL\}^{''} (\tbgam) \right]^{-1}\ba$.

Since for any $t \in [0,\tau]$, $\htheta_j(t)  =( \be_j\otimes\bB(t))^\top\hat{\bgam}$, then let $\ba = \be_j\otimes\bB(t)$, we have for any $t\in [0,\tau]$,
\begin{equation*}
\frac{\hat{\theta}_j(t)-\theta_j(t)}{\sigma_{nj}(t)} \rightarrow_d N(0,1),
\end{equation*}
where $\sigl^2(t) = n \{ \be_j\otimes\bB(t)\}^\top  \left[ -\{\rm PL\}^{''} (\bgam^*) \right]^{-1}\Sigma(\tbgam,1)\left[ -\{\rm PL\}^{''} (\bgam^*) \right]^{-1} \{ \be_j\otimes\bB(t)\}$. 

\yl{Finally, denote by $\hat{\sigma}^2_{nj}(t) = n \{ \be_j\otimes\bB(t)\}^\top  \left[ -\{\rm PL\}^{''} (\hat{\bgam}) \right]^{-1}\Sigma(\hat{\bgam},1)\left[ -\{\rm PL\}^{''} (\hat{\bgam}) \right]^{-1} \{ \be_j\otimes\bB(t)\}$. As 
$||\hat{\bgam} -\tbgam||_2 \rightarrow_p 0$, it follows that
$ \hat{\sigma}^2_{nj}(t)/ \sigl^2(t) \rightarrow_p 1 $  for $t>0$. Hence, by the Slutsky theorem,
\begin{equation*}
\frac{\hat{\theta}_j(t)-\theta_j(t)}{\hat{\sigma}_{nj}(t)} \rightarrow_d N(0,1),
\end{equation*}
which justifies the use of  $\hat{\sigma}_{nj}(t)$ as a consistent estimate of the variance.
}
 $\blacksquare$

\bibliographystyle{spbasic}
\bibliography{ch3ref}
\end{document}